\documentclass[reprint,floatfix,aps,pre,longbibliography]{revtex4-1}
\usepackage{amsmath,amssymb}
\usepackage{dsfont}
\usepackage{float}
\usepackage[colorlinks=true,linkcolor=black,citecolor=black,filecolor=black,urlcolor=black,breaklinks=true]{hyperref}
\usepackage[capitalize]{cleveref}
\usepackage{physics}
\usepackage{enumerate}
\usepackage{tikz} 
\usepackage{mathtools}
\usepackage{xcolor}
\usepackage{lmodern}

\newcommand{\cald}{\mathcal{D}} 
\newcommand{\call}{\mathcal{L}}

\newcommand{\m}{\mathrm}

\newcommand{\I}{\m{i}}
\DeclarePairedDelimiter\uketflip{\lvert}{\rangle\negthinspace\negthinspace\rangle}

\DeclarePairedDelimiterX\udyadflip[2]{\lvert}{\rvert}{#1 \delimsize\rangle\negthinspace\negthinspace\rangle \negthinspace\langle\negthinspace\negthinspace\langle #2}
\DeclarePairedDelimiterX\ubraketflip[2]{\langle\negthinspace\negthinspace\langle}{\rangle\negthinspace\negthinspace\rangle}{#1 \delimsize\vert #2}

\newcommand{\ketflip}[1]{\mathop{\uketflip{#1}}}

\newcommand{\braketflip}[2]{\mathop{\ubraketflip{#1}{#2}}}

\usepackage[colorlinks=true,linkcolor=black,citecolor=black,filecolor=black,urlcolor=black,breaklinks=true]{hyperref}
\usepackage[capitalize]{cleveref}

\begin{document}

\title{Synchronization in the quantum regime}
\author{Finn Schmolke}
\author{Eric Lutz}
\affiliation{Institute for Theoretical Physics I, University of Stuttgart, D-70550 Stuttgart, Germany}

\begin{abstract}
Can synchronization---the widespread spontaneous emergence of coordinated dynamics in classical nonlinear systems---also occur in the quantum regime? This question has recently sparked intense research into collective behavior and temporal self-organization in quantum systems. Typical quantum features such as the linearity of time evolution and the presence of quantum noise seem to hinder the appearance of synchrony at the microscopic level. At the same time, quantum coherence and quantum correlations may provide novel mechanisms for enhancing synchronization beyond its classical counterpart. We here survey recent theoretical and experimental advances in quantum synchronization, ranging from the characterization of synchronous oscillations and genuinely nonclassical forms of synchrony to many-body synchronization on quantum networks.
\end{abstract}

\maketitle

\section{Introduction}
\label{sec:q-sync}

Synchronization is a ubiquitous phenomenon in science and technology. Broadly defined as the adjustment of oscillatory systems to a common rhythm, it gives rise to phase- or frequency-locked dynamics, thus providing a universal mechanism for understanding collective motion and self-organization \cite{Kuramoto1984,Blekman1988,Mosekilde2002,Pikovsky2003,Strogatz2003,Acebron2005,Anishchenko2007,osi07,Balanov2009,Boccaletti2018}. Synchronization occurs across a wide range of physical, biological, and chemical systems, from coupled lasers to cardiac pacemaker cells \cite{Kuramoto1984,Blekman1988,Mosekilde2002,Pikovsky2003,Strogatz2003,Acebron2005,Anishchenko2007,osi07,Balanov2009,Boccaletti2018}. It has also found numerous applications in engineering, including the design of practical communication and networking systems \cite{ser09,lin17,Choi2017}.
Since its first observation by Huygens in 1665, the study of synchronized oscillations has developed into a mature  theoretical framework for classical nonlinear systems \cite{Kuramoto1984,Blekman1988,Mosekilde2002,Pikovsky2003,Strogatz2003,Acebron2005,Anishchenko2007,osi07,Balanov2009,Boccaletti2018}. Three principal forms of synchronization are commonly distinguished. First, entrainment occurs when an oscillator  adjusts its rhythm to that of an external signal, as in the synchronization of circadian processes to the light-dark cycle \cite{liu97}. Second, mutual synchronization arises when weakly coupled oscillators spontaneously swing in unison, as exemplified by metronomes on a shared support \cite{pan02}. Third, noise-induced synchronization emerges when stochastic fluctuations promote coherent behavior, such as the synchronized firing of neurons driven by common noise  \cite{nei02}.

With the advent of nanotechnology \cite{poo03} and quantum information science \cite{Nielsen2010}, research on synchronization  has lately transitioned to the quantum domain \cite{Goychuk2006,Zhirov2006,Zhirov2008}, where inherently nonclassical phenomena, such as quantum superposition and coherence \cite{arn14,str17}, as well as stronger-than-classical correlations \cite{hor09,mod12}, play a central role. Extending the concept of synchronization to quantum systems is, however, far from straightforward. The fundamental principles of quantum mechanics pose several challenges to the direct generalization of classical synchronization theory. First, the Heisenberg uncertainty principle undermines the classical phase-space description commonly used to characterize dynamical evolution, replacing it with a quasiprobability phase-space representation that may exhibit negative probabilities \cite{hil84}. Second, the linearity of quantum dynamics appears to preclude the existence of limit cycles, which constitute the cornerstone of classical synchronization phenomena \cite{Strogatz2024}. Moreover, the notion of phase itself is subtle in quantum mechanics, as a Hermitian phase operator associated with a well-defined phase observable does not generally exist \cite{lyn95}.
Despite these conceptual obstacles, remarkable  advances over the past decade have provided ways to overcome many of these difficulties, enabling the investigation of synchronization in atomic-scale systems and revealing novel forms of genuinely quantum synchrony.

We here review recent theoretical and experimental progress in quantum synchronization. We begin by introducing two complementary frameworks that have been developed to describe quantum synchrony: (i) a semiclassical approach based on the concept of quantum limit cycles, and (ii) a dynamical approach centered on correlated oscillations of local observables. These frameworks have been widely employed to characterize quantum generalizations of entrainment, mutual synchronization, and noise-induced synchronization. We also discuss the connection between quantum synchronization and quantum time crystals, which provide an alternative route to sustained synchronized oscillations.
Furthermore, we highlight several hallmark signatures of quantum synchrony, including enhanced synchronization of far-detuned oscillators, suppression of synchronization between resonant oscillators, entangled synchronous dynamics, and the phenomenon of synchronization blockade. Finally, we survey key experimental investigations of quantum synchronization and conclude with an outlook on many-body synchronization in quantum networks.

\begin{figure*}[t]
\centering
 \begin{tikzpicture}
 \node (a) [label={[label distance=-.34 cm]136: (a)}] at (0,0) {\includegraphics{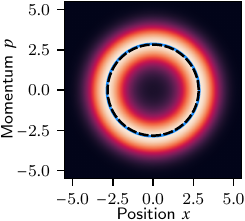}};
 \node (b) [label={[label distance=-.34 cm]136: (b)}] at (6,0) {\includegraphics{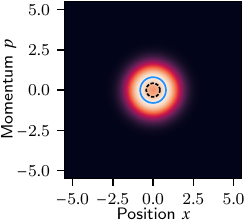}};
 \end{tikzpicture}
 \caption{Semiclassical limit cycles and Wigner distributions of the unique stationary state of the Rayleigh-van der Pol oscillator. The Wigner distributions ${W^\m{s}(x,p)}$ are computed from the stationary state of the Lindblad master equation, Eq.~\eqref{eq:quantum-vdP}. (a) Near the semiclassical regime with $\gamma^\m{p}_1 = 8\gamma_2$, the classical radius (black dashed) and the actual radius of the quantum limit cycle (blue solid) agree almost perfectly. (b) Near the quantum regime with $\gamma^\m{p}_1 = 0.18\gamma_2$, the radius of the quantum limit cycle saturates at $r_\m{q} = 1/\sqrt{2}$ due to energy quantization, while the classical radius vanishes in the limit $\gamma_2 \to \infty$.}
 \label{fig:rvdp}
\end{figure*}

\section{Semiclassical approach and quantum limit cycles}
\label{sec:quantum-limit cycles}
Classical synchronization is rooted in the concept of stable limit cycles, isolated closed orbits in the phase space of nonlinear dynamical systems, that give rise to persistent, self-sustained oscillations \cite{Kuramoto1984,Blekman1988,Mosekilde2002,Pikovsky2003,Strogatz2003,Acebron2005,Anishchenko2007,osi07,Balanov2009,Boccaletti2018}. In 2013, this framework was extended to the quantum regime through the study of a quantized version of the van der Pol oscillator \cite{Lee2013,Walter2014}, a paradigmatic system featuring nonlinear friction and serving as a cornerstone of classical synchronization theory \cite{jen13}. In dimensionless form, a broad class of classical limit-cycle oscillators can be expressed as \cite{Strogatz2024}
\begin{equation}
  \ddot{x} + x - \epsilon \dot{x} + h(x,\dot{x}) = 0,
  \label{eq:limit cycle-family}
\end{equation}
where the first three terms describe linearly damped harmonic motion.
The nonlinear damping contribution, $h(x,\dot{x})$, stabilizes an attractive, isolated phase space trajectory, with a frequency close to one if $\epsilon$ is small.
This includes the Rayleigh, $h(x,\dot{x}) \propto \dot{x}^3$ \cite{Rayleigh1883}, and van der Pol, $h(x,\dot{x}) \propto x^2\dot{x}$ \cite{vanderPol1927}, oscillators.
For weak nonlinearities, the time evolution of the system is well described by a slowly varying quasi-harmonic solution \cite{Kuramoto1984,Blekman1988,Mosekilde2002,Pikovsky2003}
\begin{equation}
  x(t) = \frac{1}{2}\left(\alpha(t)e^{\I t} + \m{c.c}\right)
\end{equation}
with a complex time-dependent amplitude $\alpha(t)$ that satisfies
the Stuart--Landau equation \cite{Kuramoto1984,Blekman1988,Mosekilde2002,Pikovsky2003}
\begin{equation}
  \dot{\alpha} = -\I \alpha + \frac{\epsilon}{2}\left(1-\lambda\frac{2|\alpha|^2}{A^2}\right)\alpha.
  \label{eq:Stuart-Landau}
\end{equation}
The asymptotic, limiting amplitude $A$ depends on the specific form of the nonlinear term $h(x,\dot{x})$\cite{Kuramoto1984,Blekman1988,Mosekilde2002,Pikovsky2003}.

An adequate quantum-mechanical generalization of such nonlinear oscillators  requires on open system approach with essentially three different processes \cite{Lee2014,Lorch2016,Lorch2017,Sonar2018,Mok2020,Arosh2021,Nadolny2023,Chia2020,Chia2025,Kehrer2025}:   a single excitation pump with rate $\gamma^\m{p}_1$, a single excitation loss with rate   $\gamma_1$, which describes linear damping, and  a two-excitation loss with rate $\gamma_2$, which corresponds to nonlinear damping.
 Accordingly, the dynamics of the  density operator $\rho$ of the quantum system can be conveniently described with  a master equation  of the Lindblad form \cite{Lee2014,Lorch2016,Lorch2017,Sonar2018,Mok2020,Arosh2021,Nadolny2023,Chia2020,Chia2025,Kehrer2025} (with the Planck constant $\hbar =1$)
\begin{equation}
 \!\! \dot{\rho} 
  = \call \rho 
  = -\I [H,\rho] + \gamma^\m{p}_1 \cald[a^\dag]\rho + \gamma_1 \cald[a]\rho + \gamma_2 \cald[a^2]\rho,
  \label{eq:quantum-vdP}
\end{equation}
where $H = a^\dag a$ is the Hamiltonian of the harmonic oscillator with ladder operators $(a, a^\dagger) $ and $\cald[X]\cdot = X \cdot X^\dag - \{X^\dag X,\cdot\}/2$ is the usual nonunitary dissipator \cite{Breuer2007}; the operator $\call$ further denotes the Lindblad superoperator.
The corresponding equation of motion for the dimensionless expectation  of the annihilation operator
\begin{equation}
  \dv{t} \langle a \rangle = -\I \langle a \rangle + \frac{\gamma_1^\m{p}-\gamma_1}{2} \langle a \rangle - \gamma_2 \langle a^\dagger aa\rangle,
  \label{eq:quantum-Stuart-Landau}
\end{equation}
though linear, is reminiscent of the classical Stuart--Landau equation \eqref{eq:Stuart-Landau}, with $(\gamma^\m{p}_1-\gamma_1) = \epsilon$
 \cite{Arosh2021}.
By  taking the expectation values above with respect to a coherent state $\ket{\alpha}$, where $\langle a \rangle = \alpha$ as well as $\langle a^\dag aa \rangle = \alpha|\alpha|^2$, one reproduces \cref{eq:Stuart-Landau} for the coherent field amplitude $\alpha$.
This approach corresponds to a mean-field decoupling where all expectation values factorize.

The Lindblad equation \eqref{eq:quantum-vdP} for the quantum van der Pol oscillator reaches a unique stationary state, if both single-particle gain and two-particle loss terms are present.
The mean occupation number in the steady-state regime is then proportional to the ratio of the corresponding rates, $\langle a^\dag a\rangle^\m{s} \propto \epsilon/\gamma_2$, corresponding to the ratio of  net  energy input and  energy  output.
Large values of the pump rate $\gamma^\m{p}_1$ result in the population of highly excited states, associated with the semiclassical limit where quantum fluctuations become negligible, while large two-particle loss rate $\gamma_2$ corresponds to the deep quantum regime, where only the lowest few Fock states of the oscillator are populated, and quantum fluctuations become dominant \cite{Lee2013,Walter2014}.

The properties  of the quantum van der Pol oscillator can be analyzed analogously to those of the classical oscillator with  the help of the Wigner  function \cite{Lee2014,Lorch2016,Lorch2017,Sonar2018,Mok2020,Arosh2021,Nadolny2023,Chia2020,Chia2025,Kehrer2025}\begin{equation}
\label{6}
  W(x,p) = \frac{1}{\pi} \int_{-\infty}^\infty \dd{y} \bra{x-y}\rho \ket{x+y} e^{2\I py},
\end{equation}
which provides the quantum phase portrait for the noncommuting observables position and momentum \cite{hil84}.
\Cref{fig:rvdp} shows the Wigner distribution of the quantum limit cycle oscillator described by the stationary state of the Lindblad equation \eqref{eq:quantum-vdP} without linear damping $\gamma_1 = 0$.
Similarly to a classical harmonic oscillator, the distribution of the steady state, $W^\m{s}(x,p)$, exhibits a ring shape with a radius that depends on the gain and loss rates. However, 
due to the Heisenberg uncertainty principle, what used to be points in classical phase space, are now broadened to distributions in the quantum-mechanical quasi phase portrait. The classical case is recovered in the limit of large steady-state mode occupation, 
$\gamma^\m{p}_1/\gamma_2 \gg 1$, where quantum effects become  negligible, and the Wigner function approaches the phase-space distribution of the classical limit cycle oscillator (\cref{fig:rvdp}a).
Energy quantization becomes relevant in the opposite regime, $\gamma^\m{p}_1/\gamma_2 \ll 1$, where a first excited mode cannot be depleted by the two-excitation loss process. As a result, the ground and first excited states are always occupied, and the quantum radius approaches a nonzero value, $r_q = 1/\sqrt{2}$, in the limit  $\gamma_2 \to \infty$, in contrast to the classical radius (\cref{fig:rvdp}b).

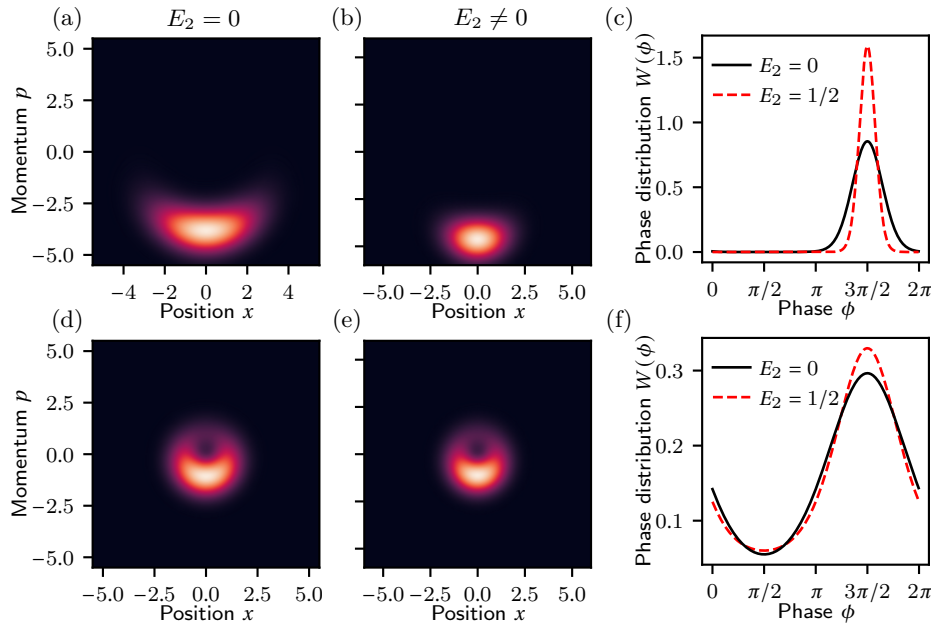
\begin{figure*}[t]
  \begin{tikzpicture}
    \node (a) at (0,0) {\includegraphics{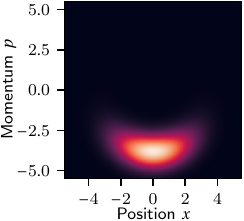}};
    \node (b) at (4,0) {\includegraphics{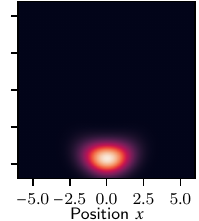}};
    \node (c) at (8,0) {\includegraphics{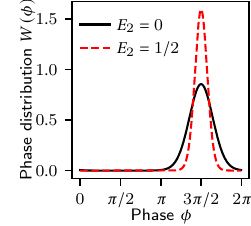}};

    \node (d) at (0,-4) {\includegraphics{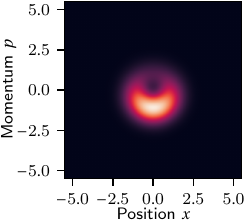}};
    \node (e) at (4,-4) {\includegraphics{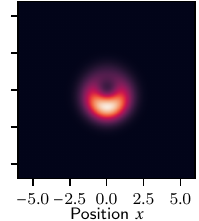}};
    \node (f) at (8,-4) {\includegraphics{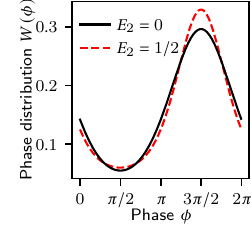}};
    \draw (0.5,2.1) node {\small $E_2=0$};
    \draw (4.3,2.1) node {\small $E_2\neq 0$};

    \draw (-1.3,2.1) node {(a)};
    \draw (2.4,2.1) node {(b)};
    \draw (6,2.1) node {(c)};
    \draw (-1.3,-1.9) node {(d)};
    \draw (2.4,-1.9) node {(e)};
    \draw (6,-1.9) node {(f)};
  \end{tikzpicture}

 \caption{Entrainment of the quantum limit cycle oscillator, \cref{eq:rvdp-full}, tuned to an external drive ($\Delta = 0$) with and without squeezing. (a) Wigner distribution in the near classical regime, $\gamma^\m{p}_1 = 8\gamma_2$, in the co-rotating frame. 
 The drive ($E=1$) increases the mode occupation and confines the distribution in phase space.
 (b) Squeezing ($\theta=\pi/2$ in \cref{eq:rvdp-full}) injects more energy and concentrates the distribution in one direction while broadening it in the perpendicular direction. 
 (c) The phase distribution $W(\phi)$ \eqref{eq:phase-distribution} is peaked around the preferred phase and is narrowed in the presence of squeezing, leading to improved phase locking.
 In the near quantum regime, $\gamma^\m{p}_1=0.18\gamma_2$, quantum noise hinders entrainment ($E=1$) and (d),(e) the Wigner function remains delocalized.
 (f) While $W(\phi)$ still indicates a preferred phase, squeezing yields a slight improvement.
 }
 \label{fig:entrainment}
\end{figure*}

Although it is tempting to consider the stationary Wigner distribution as a quantum limit cycle based on the behavior of the phase-space trajectories in the classical limit, in the deep quantum regime this interpretation is no longer immediate because phase-space trajectories are no longer available.
To actually observe the oscillating dynamics of a system, quantum measurements have to be performed.
However, due to measurement backaction, observing a quantum system  strongly affects its dynamics \cite{Barchielli2009,Wiseman2009,Jacobs2014,Jordan2024}. 
A particularly well-suited method is indirect continuous monitoring that unravels the stationary state into stochastic quantum trajectories with minimal disturbance \cite{Barchielli2009,Wiseman2009,Jacobs2014,Jordan2024}.
Based on this trajectory picture, a quantum theory of phase reduction, that
allows one to identify the underlying limit cycle evolution, has been recently introduced  in Refs.~\cite{Setoyama2024,Setoyama2025}.
For the van der Pol model, the corresponding limit cycle radius coincides exactly with the radius of the Wigner distribution; the classical theory is additionally recovered  in the appropriate limit \cite{Setoyama2024,Setoyama2025}.
In this regard, the notion of a quantum limit cycle  can be meaningfully extended  to the quantum regime.
The study of quantum limit cycles is an active and rapidly evolving  research area and many open questions are currently explored, including systems featuring multiple limit cycles and limit tori \cite{Kehrer2025,YiZhao2025,Nadolny2026,Chen2026,Hassler2026,Christiansen2026}.

Physical systems to which the   semiclassical approach to quantum synchronization has been applied to include trapped ions \cite{Lee2013,hus15} and optomechanical systems \cite{Heinrich2011,Ludwig2013,Weiss2016}, for which a truncated Wigner approximation eventually yields semiclassical equations that can be compared to the full solution of the quantum master equation. In addition, higher order and strong nonlinearities have also been  explored to study quantum analogs of relaxation oscillators with strongly nonuniform  limit cycles \cite{Chia2020,Chia2025}.
The corresponding  equations  for quantum continuous-variable oscillators with arbitrary nonlinearities obtained from a general quantization scheme aim for a faithful description of  classical nonlinear models, and reproduce phase portraits close to the ones produced by noisy classical dynamics \cite{Chia2020,Chia2025}.

The above quantum synchronization framework has also been successfully applied to quantum systems   with a finite number of energy levels such as spin systems \cite{Roulet2018a,Roulet2018,Koppenhofer2019,Lopez2020,Kehrer2024,Kato2025}.
The Wigner distribution is in this case replaced by the Husimi $Q$ function, defined as \cite{Gilmore1975}
\begin{equation}
\label{7}
  Q(\theta,\phi) = \frac{2S+1}{4\pi} \bra{\theta,\phi} \rho \ket{\theta,\phi},
\end{equation}
for spin-coherent states $\ket{\theta,\phi}$, where $S$ is the total spin \cite{Chryssomalakos2018,Kam2023}.
Although the usual semiclassical correspondence is not available, quantum limit cycles may be identified as phase symmetric stationary states of Lindblad equations and entrainment as well as synchronization can be studied in a similar fashion, while the large spin-limit is able to recover the properties of the quantum Rayleigh-van der Pol oscillator \cite{Kato2025}.

\subsection{Quantum entrainment}

The semiclassical formalism has been employed to examine both quantum entrainment and quantum mutual synchronization.
Entrainment is the unidirectional synchronization of oscillators to an external periodic signal \cite{Kuramoto1984,Blekman1988,Mosekilde2002,Pikovsky2003,Strogatz2003,Acebron2005,Anishchenko2007,osi07,Balanov2009,Boccaletti2018}.
For quantum limit cycle oscillators, entrainment is possible by introducing the action of an external drive \cite{Lee2013,Walter2014}.
 A squeezing drive can additionally be considered to potentially enhance synchronization \cite{Sonar2018}.
In the corotating reference frame, the Hamiltonian then reads
\begin{equation}
  H = \Delta a^\dagger a + \I E (a-a^\dagger) + \I \frac{E_2}{2} ({a^\dagger}^2e^{2\I \theta} - a^2e^{-2\I \theta}),
  \label{eq:rvdp-full}
\end{equation}
where $\Delta$ is the detuning between the oscillator and the drive, $E$ is the driving strength, $E_2$ is the squeezing strength and $\theta$ is the relative phase between the drives.
The semiclassical approximation for the coherent field amplitude  $\alpha$ becomes accordingly
\begin{equation}
  \dot{\alpha} = \I \Delta \alpha + \frac{\gamma^\m{p}_1-\gamma_1}{2}\alpha -\gamma_2 |\alpha|^2 \alpha - E_2 e^{\I \theta} \alpha^\ast - E.
\end{equation}
Decomposing the complex amplitude as $\alpha = r\exp(\I \phi)$, the equation of motion can be recast in polar coordinates for the radius $r$ and the phase $\phi$ to yield \cite{Walter2014,Sonar2018}
\begin{align}
 \!\!\!\!\!\! \dot{r} &= \frac{\gamma^\m{p}_1-\gamma_1}{2} r - \gamma_2 r^3 - 2E_2 r \cos(2\phi-\theta) - E \cos(\phi),\\
  \!\!\dot{\phi} &= \Delta + E_2 \sin(2\phi-\theta) + \frac{E}{r} \sin(\phi). \label{10}
\end{align}
{When the time dependence of $r$ is negligible, the} semiclassical evolution of the phase variable $\phi$ given by \eqref{10} is analogous to   the classical Adler equation, modified by the influence of the squeezing drive. The Adler equation is an effective description of a weakly forced limit cycle oscillator, reducing the full problem to a closed equation for the phase alone \cite{Nakao2016}.
\Cref{fig:entrainment} shows the Wigner distribution for the quantum oscillator subject to external driving.
The semiclassical regime is well described in terms of a Langevin equation given by the Stuart--Landau model in the presence of white noise, where the Wigner function can be approximated by a Fokker--Planck equation \cite{Lee2013}.
The drive has the effect of localizing the Wigner function in phase space, indicating phase locking at the drive's frequency, a hallmark of synchronization.
Whereas the circular shape of a free quantum limit cycle (\cref{fig:rvdp}) is characterized by a diagonal density operator in the Fock basis, the deformed phase portrait of an entrained quantum oscillator (\cref{fig:entrainment}) features off-diagonal matrix elements, signaling the presence of quantum coherence \cite{Arosh2021}.
The phase distribution
\begin{equation}
  W(\phi) = \int \dd{r} r W(r,\phi),
  \label{eq:phase-distribution}
\end{equation}
can be readily obtained by marginalizing the Wigner function \eqref{6} in polar coordinates.
In the near classical regime (\cref{fig:entrainment}a--c), the phase distribution is approximately normal and becomes increasingly concentrated around a preferred phase for larger driving strength $E$, revealing a clear phase preference of the oscillator.
Moreover, squeezing perpendicular to the drive with $\theta=\pi/2$, reduces the uncertainty in the phase at the expense of increasing the uncertainty in the amplitude.
At finite energies, the quantum oscillator can never be perfectly entrained due to the quantum noise introduced by the Heisenberg uncertainty principle.
At low energies (\cref{fig:entrainment}d-f), the Wigner function and the phase distribution are considerably broadened, and the oscillator is less easily entrained.
While the enhancing effect of squeezing is still present, it is less pronounced compared to the semiclassical regime.
Quantization prevents a complete loss of entrainment in the deep quantum regime that occurs in the semiclassical model.

\subsection{Mutual quantum synchronization}
\label{sec:mutual-sync}

Another paradigm of classical synchronization is mutual synchronization of several oscillators \cite{Kuramoto1984,Blekman1988,Mosekilde2002,Pikovsky2003,Strogatz2003,Acebron2005,Anishchenko2007,osi07,Balanov2009,Boccaletti2018}.
Multiple quantum oscillators can be synchronized to each other by employing either coherent (reactive) coupling,  where direct interactions between the  $N$ oscillators are incorporated in an interaction Hamiltonian $H_\m{int} = (V/N)\sum_{m<n}(a^\dag_m a_n+\m{h.c.})$, or, through incoherent (dissipative) processes, added to the dissipator with $(V/N)\sum_{m<n}\cald[a_m-a_n]$, where $a_m$ denotes the annihilation operator of the $m$-th oscillator and $V$ is the coupling strength \cite{Lee2013,Lee2014,Walter2014,Walter2015}.
The framework developed for quantum entrainment is naturally extended to mutual synchronization.
The phase space description is still available by considering the joint Wigner function $W(\vb{x},\vb{p})$, with position and momentum coordinates $\vb{x}=(x_1,\ldots,x_N)$ and $\vb{p}=(p_1,\ldots,p_N)$.
Alternatively, one may marginalize the Wigner function to obtain the distribution of the phase difference, $W^\m{s}(\phi_1-\phi_2)$, as an indicator of phase locking.
Similar to the classical case, there are two maxima corresponding to in-phase ($\phi_1-\phi_2=0$) and anti-phase ($\phi_1-\phi_2=\pi$) synchronization.
In the quantum model, these maxima appear if the steady state has coherences of energy levels beyond the first excited state.
{They} become washed out at lower energies and vanish in the quantum limit ($\gamma_2/\gamma_1^p \to \infty$).

For many oscillators, and in analogy to the classical Kuramoto model \cite{Acebron2005}, an order parameter $R = (1/N)\sum_n \langle a_n \rangle$ can be defined, which measures the coherence of the oscillators, where $R>0$ indicates the presence of synchronization \cite{Lee2013,Lee2014,Tilley2018}.
Reactively coupled quantum oscillators at low energies are more easily synchronized than the corresponding classical model subject to white noise; they already exhibit a phase transition into the synchronized phase for smaller values of the coupling strength $V$ \cite{Lee2013,Tilley2018}. Dissipative coupling leads to a synchronized phase even in the presence of frequency disorder, but here, the required critical coupling strength is much larger compared to the classical counterpart \cite{Lee2014}.

\section{Dynamical approach and synchronization transitions}

Phase-space quantum synchronization is able to replicate many  features of the classical theory, while adding new insight into the nature of synchronous behavior in the deep quantum regime. Starting in 2010, a complementary dynamical approach, where synchronization is induced by dissipative processes that drive a transition from unsynchronized to synchronized evolution of local observables,  has been developed \cite{Orth2010,Giorgi2012,Giorgi2013,Galve2017,Bellomo2017,Karpat2019,Cabot2019,Giorgi2019,Buca2022,Schmolke2022,Schmolke2024}. In this framework, quantum synchronization can be explored  in systems without a classical analog.
Synchrony between two local observables $\langle O_m\rangle$ and $\langle O_n\rangle$  of  a system can often be measured by the Pearson correlation coefficient \cite{Barlow1993}
\begin{equation}
  C_{mn} = \frac{\m{Cov}(\langle O_m\rangle,\langle O_n\rangle)}{\sqrt{\m{Var}(\langle O_m\rangle)\m{Var}(\langle O_m\rangle)}},
  \label{eq:Pearson}
\end{equation}
that quantifies their linear temporal correlation.
Any nontrivial time periodic pair of observables with $C_{mn}=\pm1$ is identically (anti)synchronized, while $C_{mn}=0$ indicates no linear correlation.

\subsection{Stable synchronization}

In closed quantum systems, identically oscillating subsystems can be trivially realized, simply by selecting an appropriate initial condition.
However, the required nonlocal control may be unfeasible in an actual experimental setting.
An additional requirement is therefore that synchronization should emerge spontaneously, be robust against perturbations and occur generically for a wide range of initial states.
The problem of synchronization then boils down to establishing a unique invariant manifold that attracts (almost) all trajectories in Hilbert space towards periodic stable orbits.
It is  often convenient to adopt a master equation in Lindblad form \cite{Stefanini2025}
\begin{equation}
  \dot{\rho} 
  = \call \rho
  = -\I[H_0 + H_\m{int},\rho] + \sum_\mu \left(L_\mu \rho L^\dag_\mu - \frac{1}{2}\left\{L^\dag_\mu L_\mu,\rho\right\}\right),
\end{equation}
with Lindblad jump operators $L_\mu$, to describe the time evolution of the system.
The synchronization properties follow from a spectral analysis of the generator $\call$, which generally features complex eigenvalues $\Lambda_k$.
If $\call$ is diagonalizable, using vectorization that enables the description of a quantum system in Liouville space with a density vector $ \ketflip{\rho(t)}$ instead of the density matrix $\rho(t)$ in  the usual state space \cite{Gyamfi2020}, the dynamics of the state can be formally decomposed according to \cite{Gyamfi2020}
\begin{equation}
  \ketflip{\rho(t)} = \sum_k e^{\Lambda_k t} \braketflip{\tau_k}{\rho(0)} \ketflip{\sigma_k},
\end{equation}
with biorthonormal left and right eigenvectors $\braketflip{\tau_j}{\sigma_k}=\delta_{jk}$ and complex eigenvalues $\Lambda_k$, satisfying $\m{Re}[\Lambda_k] \le 0$.
In the theory of open quantum systems  \cite{Breuer2007}, periodic orbits necessarily correspond to purely imaginary eigenvalues, $\Lambda_k = \I \lambda_k$, $\lambda_k \in \mathbb{R}$, and are closely related to steady-state degeneracy and information preserving structures, such as noiseless subsystems and decoherence-free subspaces that are decoupled from the surroundings and thus support unitary evolution \cite{Lidar1998,Knill2000,Baumgartner2008,Baumgartner2012,Blume2010}.
Let the eigenvalues be arranged by decreasing real part ($\m{Re}[\Lambda_k] \ge \m{Re}[\Lambda_{k+1}]$).
Up to exponentially small corrections, the asymptotic evolution of any observable $O$ is then coherent, i.e.
\begin{equation}
  \langle O(t)\rangle \sim \sum_k e^{\I\lambda_k t} c_k O_k + \mathcal{O}(e^{-\gamma t}),
  \label{eq:coherent-observable}
\end{equation}
with coefficients $c_k = \braketflip{\tau_k}{O}$ and $O_k = \braketflip{O}{\sigma_k}$.
The synchronization time is determined by the spectral gap $\gamma = -\m{Re}[\Lambda_1]$ which corresponds to the decay rate of the slowest mode.
While these considerations are sufficient for coherent evolution, they are not yet enough for synchronized dynamics.
To guarantee synchronization between sites $m$ and $n$, one additionally has to demand permutational invariance of imaginary eigenmodes \cite{Buca2022}.
Still, if the eigenvalues are too chaotic, they might destructively interfere and lead to equilibration of local observables as for instance predicted by the eigenstate thermalization hypothesis \cite{Deutsch1991,Srednicki1999,DAlessio2016,Deutsch2018}.

A sufficient criterion for imaginary modes is the existence of one or several strong dynamical symmetries with operators $A_k$ that are eigenoperators of the unitary evolution and strong symmetries of all Lindblad operators, satisfying the conditions \cite{Buca2022,Baumgartner2008,Albert2018}
\begin{equation}
  [H,A_k] = \lambda_k A_k, \qquad [L_\mu,A_k] = [L^\dag_\mu,A_k] = 0, \ \forall \mu.
\end{equation}
For nonzero $\Lambda_k$ every $A_k$ is associated with a finite ladder of equally spaced imaginary eigenvalues $\Lambda_k = -\I \lambda_k n$, $n=1,2,\ldots$.
In this case, the operator $A_k$ must be both traceless and singular.
There may be many eigenvalue towers or only a single $A$, generating a single imaginary eigenvalue.
Similar structures also appear in closed quantum many-body scarred systems, where the generating algebra gives rise to regular islands embedded in otherwise chaotic systems \cite{Serbyn2021}.
A necessary prerequisite for dynamical synchronization is steady-state degeneracy of the quantum master equation.
Corresponding research is thus interrelated to engineered dissipation and ergodicity breaking in few and many-body quantum systems {\cite{Albert2018,Harrington2022,Buca2022,Schmolke2024}}.
These requirements contrast with the quantized nonlinear oscillators models of the previous section, most of which exhibit a unique asymptotic nonequilibrium state.

\begin{figure}[t]
  \centering
  \begin{tikzpicture}
  \node (a) at (0,0) {\includegraphics{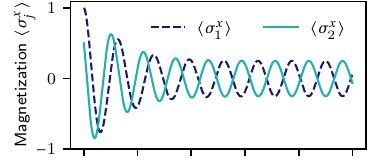}};
  \node (b) at (0,-3.1) {\includegraphics{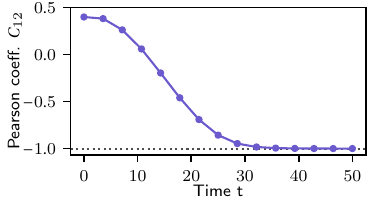}};
  \draw (-3.5,1.5) node {(a)};
  \draw (-3.5,-1.4) node {(b)};
  \end{tikzpicture}
  \caption{Dynamical synchronization. (a) Stable antisynchronization of the $x$-magnetizations of two qubits undergoing collective spontaneous emission into a common zero-temperature bath. Undesired eigenmodes are exponentially suppressed until only the permutationally invariant decoherence-free mode survives, protecting the asymptotic coherent evolution. (b) Pearson correlation coefficient, \cref{eq:Pearson}, indicating perfect linear anticorrelation between $\langle \sigma^x_1 \rangle$ and $\langle \sigma^x_2 \rangle$.}
  \label{fig:stable-sync}
\end{figure}

Once the stable synchronization conditions are met, the evolution will exponentially suppress all undesired eigenmodes until only the permutationally invariant, decoherence-free modes are left, protecting the asymptotic coherent dynamics. An example of such process is shown in  \cref{fig:stable-sync} for two qubits coupled to a common zero-temperature bath as discussed in Ref.~\cite{Karpat2020}. An important feature of classical synchronization is its robustness.
By design, dynamical quantum synchronization is exponentially stable against perturbations that push the state out of the invariant manifold, while tangential perturbations can at most affect the oscillation amplitudes \cite{Buca2022}.

Noise-induced synchronization can spontaneously occur in quantum systems when they are subject to classical \cite{Schmolke2022} or quantum \cite{Schmolke2024} noise. In this case, noise drives the evolution into a decoherence-free subspace of the system, leading to stable synchronized oscillations of local observables, such as distant magnetizations along a spin chain \cite{Schmolke2022,Schmolke2024}. This phenomenon extends the concept of classical noise-induced synchrony \cite{Goldobin2005,Nakao2007,Teramae2008}, which has, for example, been observed  in lasers \cite{Sunada2014}.

\subsection{Metastable synchronization}

Dynamical quantum synchronization requires exact Hilbert space symmetries.
However, in realistic settings there is typically finite leakage into the protected subspaces that eventually damps out any nonstationary behavior.
If ideal protection is not available, synchronization may still persist over a finite, but considerably long time span compared to the period of the oscillations  \cite{Giorgi2012,Giorgi2013,Galve2017,Bellomo2017,Karpat2019,Cabot2019,Giorgi2019}.
In this case of metastable synchronization, purely imaginary eigenvalues and steady-state degeneracy are entirely absent.
However, let the eigenvalues of the Lindbladian be arranged by decreasing real part
\begin{equation}
  \m{Re}[\Lambda_0] \le \m{Re}[\Lambda_1] \le \m{Re}[\Lambda_2] \le \ldots,
\end{equation}
with $\Lambda_0 = 0$, corresponding to the unique stationary state $\rho^\m{s}$.
If there exists a large separation between consecutive eigenvalues $\m{Re}[\Lambda_m]/\m{Re}[\Lambda_{m+1}] \ll 1$, this implies a timescale separation in the dynamics where the subset $\{\ketflip{\sigma_k}\}_{k=1}^{k=m}$ of modes decays significantly slower than every other mode \cite{Giorgi2012,Giorgi2013,Galve2017,Bellomo2017,Karpat2019,Cabot2019,Giorgi2019}.
Synchronization may thus appear over an extensive time period, although eventually, the system relaxes to a unique nonequilibrium stationary state.
For these oscillations to be synchronized, permutational invariance must now hold for the slowest decaying modes.
Metastable, nearly coherent, dynamics of local observables are then expected over a range of $1/\abs{\m{Re}[\Lambda_{m+1}]} \ll t \ll 1/\abs{\m{Re}[\Lambda_m]}$ \cite{Macieszczak2016} with \footnote{In general, a more careful analysis is required that takes into account both the interplay between the eigenvalues and the eigenmodes to correctly estimate the relaxation time scales. Knowledge of the real part of the spectrum $\m{Re}[\Lambda_1],\m{Re}[\Lambda_m]$ and $\m{Re}[\Lambda_{m+1}]$ may no longer provide sufficient information, if for instance the coefficients $c_k, \ (k \ge m+1)$ are anomalously large and thus modify the decay rates at intermediate times \cite{Song2019,Mori2020,Haga2021,Lee2023}.}
\begin{equation}
  \langle O(t)\rangle \sim \sum_{k=0}^m e^{\I\lambda_k t} c_k O_k + \mathcal{O}(\m{Re}[\Lambda_m] t) + \mathcal{O}(e^{\m{Re}[\Lambda_{m+1}]t}).
\end{equation}
The techniques and methods to identify such transient synchronization phenomena are hence closely related to the relaxation of open quantum systems in general and to establishing long-lived quantum coherences in particular \cite{Znidaric2015,Letscher2017,Minganti2018,Mori2020,Haga2021,Lee2023}.
Based on the spectral analysis of the Lindblad generator, a theory of metastability in quantum systems has been developed in Ref.~\cite{Macieszczak2016}.

Metastable quantum synchronization can often be understood from perturbation theoretic considerations, where a small additional control Hamiltonian or dissipative process weakly breaks a symmetry and lifts the steady state degeneracy.
Small perturbations to the Liouvillian can be accounted for by a series expansion \cite{Buca2022}
\begin{equation}
  \call(g) = \call + g \call^{(1)} + g^2 \call^{(2)} + \ldots,
\end{equation} 
in the small parameter $g$.
The eigenvalues are expanded accordingly
\begin{equation}
  \Lambda_k(g) = \Lambda_k + g\Lambda^{(1)}_k + g^2\Lambda^{(2)}_k + \ldots.
\end{equation}
Additional dissipation may (i) render previously stable synchronization metastable (if $\Lambda_k=\I \lambda_k$ and $\m{Re}[\Lambda^{(1)}_k]< 0$) or (ii) induce oscillations when there were previously none (if $\Lambda_k=0$ and $\Lambda^{(1)}_k=\I \lambda_k$) \cite{Buca2022}.
Case (i) represents the generic behavior.
This happens when a perturbation weakly breaks a strong symmetry, thereby introducing finite leakage into previously protected subspaces through which dissipation can eventually reach the entire asymptotic state space.
{Case} (ii) is realized, for instance, by strong measurements in the Zeno regime that can create quasi-decoherence-free manifolds with a lifetime inversely proportional to the measurement strength \cite{Facchi2008,Zanardi2014,Popkov2018,Burgarth2020,Popkov2021}.

Metastable synchronization has been analyzed in numerous models with collective or local dissipation \cite{Giorgi2012,Giorgi2013,Galve2017,Bellomo2017,Karpat2019,Cabot2019,Giorgi2019}, for example,
 in relation to superradiance, in a dimer atomic lattice with both staggered level spacings and loss, and in a collisional model \cite{Giorgi2012,Giorgi2013,Galve2017,Bellomo2017,Karpat2019,Cabot2019,Giorgi2019}.
The physical parameters not only determine the relaxation time scales but may also affect the eigenmodes.
This fact can be used to transition between in-phase and anti-phase synchronization by varying the detuning between two qubits \cite{Galve2017}.
Since metastable synchronization exists only in the pre-asymptotic regime, the stationary state has only little {significance}. 
On the other hand, {non-Markovian effects} are relevant at intermediate time scales, and thus have a much greater influence \cite{Breuer2016,deVega2017}.
Information backflow can, {for instance}, disrupt pre-existing transient synchronized modes \cite{Karpat2021,Zhou2021}.
Although further investigation is needed, one may also imagine increased stability and prolonged metastable regimes induced by non-Markovian effects.
Lastly, we point out that metastable and stable synchronization are not mutually exclusive; indeed the theory does  not forbid pre-synchronized metastable oscillations that transition into a stable, synchronized regime.

\section{Relation to quantum time crystals}

The notion of stable quantum synchronization is related to quantum time crystals \cite{Wilczek2012,Sacha2018,Khemani2019,Zaletel2023,Iemini2018,Hadjusek2022,Krishna2023,Solanki2024,Russo2025}, which provide an alternative route towards stable temporal coordination.
Quantum time crystals are characterized by a spontaneous time-symmetry breaking in the thermodynamic limit, where time-like crystalline order emerges \cite{Wilczek2012,Sacha2018,Khemani2019,Zaletel2023,Iemini2018,Hadjusek2022,Krishna2023,Solanki2024,Russo2025}.
The resulting collective motion is typically synchronized and exhibits desirable stability properties \cite{Sacha2018,Zaletel2023}. However, not all synchronized systems are time crystals.
Early studies have focused on closed systems where a periodic external drive elicits a rigid subharmonic response (discrete time crystals) \cite{Sacha2018,Khemani2019,Zaletel2023}, with subsequent generalizations that included dissipation and continuous-time symmetry breaking (dissipative continuous time crystal) \cite{Iemini2018,Hadjusek2022,Krishna2023,Solanki2024,Russo2025}.
Dynamical synchronization of dissipative time crystals without external driving was 
demonstrated in a many-body open Hubbard model \cite{Buca2019} and finite-level systems \cite{Iemini2018,Buonaiuto2021,Hadjusek2022,Krishna2023,Solanki2024,Russo2025}.
The stability properties are typically inherited from the effective nonlinearities that emerge in the thermodynamic limit.
Quantum correlations are hence partially neglected \cite{Sacha2018,Khemani2019,Zaletel2023,Russo2025}.
Quantum synchronization, by contrast, naturally aims at an understanding in the deep quantum regime.

\section{Quantum signatures of synchronization}

The framework of quantum synchronization not only extends conventional classical synchronization to the quantum domain, it also leads to nonintuitive genuinely quantum forms of synchrony. 
Let us, for example, consider a limit cycle, which corresponds to a stable amplitude where energy gain and dissipation are balanced.
In classical systems, when the dissipation increases, the amplitude shrinks until it overwhelms the gain completely, and no oscillations are possible anymore.
Intriguingly, this intuition generally breaks down for quantum limit cycles due to energy quantization.
Both energy loss and gain correspond to an exchange of discrete quanta, and to stabilize a quantum limit cycle the number of excitations of each process typically does not match, such that even at arbitrarily strong dissipation, a finite amount of levels remains excited.
Quantum entrainment hence becomes possible in regions where it is classically forbidden \cite{Lorch2016,Lorch2017}.
On the other hand, quantum synchronization defies our classical intuition and may be absent when one would expect it to be most prominent, namely when the oscillators have the same natural frequency \cite{Lorch2016,Lorch2017}.
In general, the exchange of discrete energy quanta must be resonant, which becomes more relevant at lower energies where energy quantization may actually obstruct mutual synchronization.
When considering an anharmonic contribution to the Hamiltonian, resulting for instance from a Kerr-nonlinearity $\propto K (a^\dag a)^2$, the energy levels are no longer equally spaced, and the resonance condition is offset; synchronization is hence more pronounced for detuned oscillators \cite{Lorch2016,Lorch2017}.
These synchronization blockades seem to be a generic feature of quantum synchrony and can have different physical origins \cite{Lorch2016,Lorch2017,Roulet2018a,Roulet2018,Solanki2023,Lopez2020,Dai2026}.
Forbidden parameter regions can, for instance appear as a consequence of unfavorable symmetries, where coherences in the steady state destructively interfere to prevent synchronization \cite{Roulet2018a,Roulet2018,Koppenhofer2019,Solanki2023,Kehrer2024}.
These synchronization blockades are only present in the full quantum-mechanical description of synchronization.

\begin{figure}[t]
  \centering
    {\includegraphics[scale=.385]{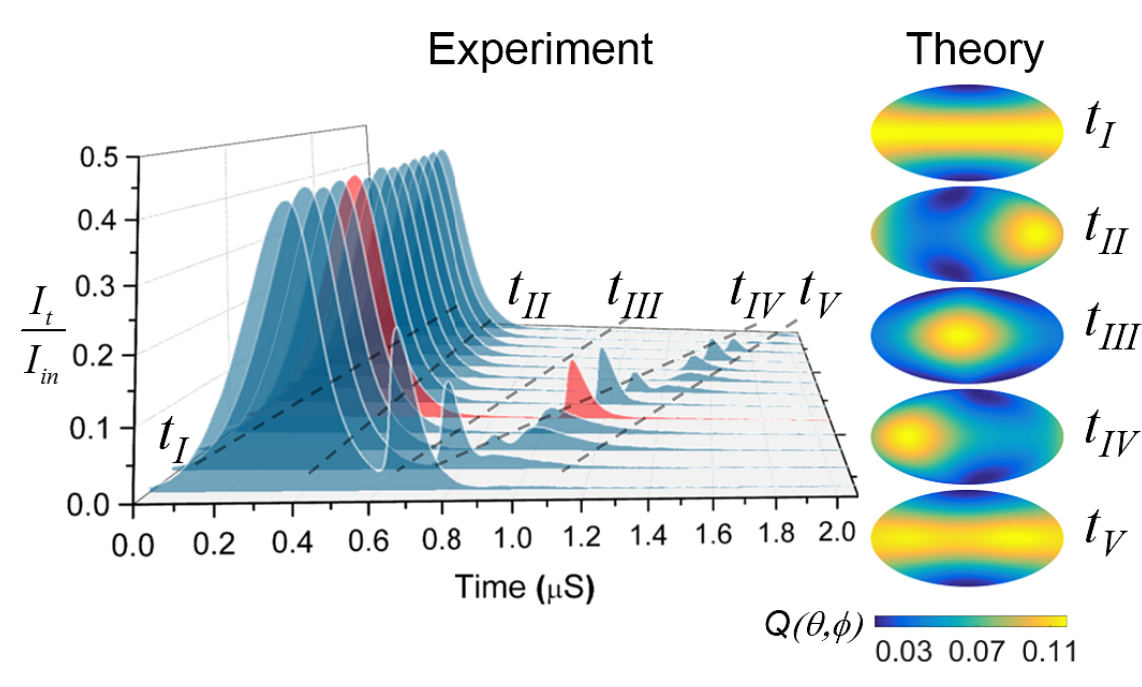}}
  \caption{{Entrainment of spin-1 $^{87}$Rb atoms. Relative intensities $I_\m{t}/I_\m{in}$ of transmitted and input {signals}, as a function of time  for different values of the storage time (experiment) and corresponding Husimi $Q$ functions, Eq.~\eqref{7}, (theory). Starting from a limit cycle at time $t_\m{I}$, an external drive entrains the atoms at times $\tau \in [t_\m{II},t_\m{IV}]$, leading to a localized and precessing Husimi $Q$ function (taken from Ref.~\cite{Laskar2020}).}}
  \label{fig:Laskar}
\end{figure}

Stronger-than-classical correlations, such as entanglement, can be achieved in the synchronization of   quantum oscillators in the semiclassical approach \cite{Roulet2018,Koppenhofer2019}. 
Although in the deep quantum regime phase locking cannot be perfect, all-to-all coupled oscillators become entangled beyond a critical coupling $V_\m{c}$, when their frequencies match \cite{Lee2014}.
For nonzero detuning, stronger coupling is required to entangle and at large frequency mismatch, the phase boundary becomes proportional to the detuning $V_\m{c} \sim |\Delta|/2$.
Due to its formal resemblance with the classical Arnold tongue, the corresponding phase diagram is  referred to as the entanglement tongue of quantum synchronization \cite{Lee2014,Roulet2018,Ameri2015,Koppenhofer2019,Kehrer2024}.
The entanglement tongue has also been shown to provide a witness for synchronization in two coupled spin-1 oscillators \cite{Roulet2018}.
Although coherences are necessary for sustained oscillations, quantum synchronization does not always imply the presence of entangled oscillations \cite{Giorgi2012,Giorgi2013,Mari2013,Lee2014,Ameri2015,Roulet2018,Giorgi2019,Witthaut2017}.

Quantum correlations have also been shown to appear in the dynamical synchronization framework \cite{Giorgi2012,Manzano2013,Galve2017,Cabot2018,Tindall2020,Buca2022,Schmolke2022,Tong2025,Cakmak2026}, which exhibits quite complementary quantum features compared to the semiclassical approach.
Synchronized oscillations can here only be sustained by surviving coherences, and originate from a competition of entangling unitary dynamics and disentangling decay, dissipation and dephasing.
The surviving eigenmode's ability to hold quantum correlations generally determines the amount of mutual information, entanglement or discord that is expected to be built up \cite{Giorgi2012,Manzano2013,Galve2017,Cabot2018,Tindall2020,Buca2022,Schmolke2022,Tong2025,Cakmak2026}.
Synchronization can hence serve as a means to correlate and entangle different subsystems.
In fact, information-theoretic quantifiers,  such as mutual information and relative entropy, have been proposed to detect quantum synchronization \cite{Ameri2015,Jaseem2020}.

{In addition, the} study of dynamical quantum synchronization also opens up the possibility to find new synchronization phenomena without classical correspondence.
Stable synchronization can be induced by continuous indirect observation of a quantum system where the measurement backaction drives the whole system into a protected subspace such that it entirely escapes the influence of the environment \cite{Schmolke2024}. Synchronization may appear at the trajectory level while being absent at the ensemble level--and vice versa. In addition, individual synchronized trajectories at multiple frequencies are possible by engineering coherent superpositions of decoherence-free subspaces with distinct frequencies, leading to a quantum form of multiplexing \cite{Schmolke2024}.

\begin{figure}[t]
  \centering
   {\includegraphics[scale=0.367]{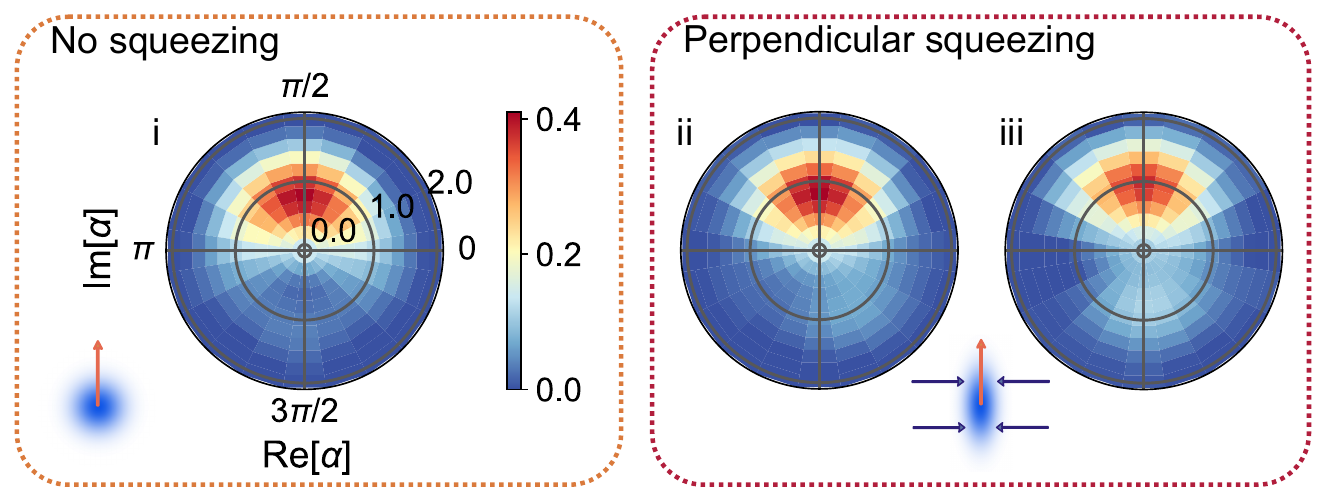}}
  \caption{{Synchronization of a quantum van der Pol oscillator realized with  a single trapped $^{40}$Ca$^{+}$ ion. An external drive localizes the Wigner function, Eq.~\eqref{6}, around $\pi/2$, revealing phase locking along  the drive (without squeezing). Squeezing perpendicular to the drive improves synchronization by reducing the effects of quantum noise (taken from Ref.~\cite{Li2025}).}}
  \label{fig:Li}
\end{figure}

\section{Experimental observations}

A growing number of experimental studies have reported  the observation of  quantum features of synchrony  on different physical platforms, ranging from cold atoms \cite{Laskar2020} and superconducting qubits \cite{Koppenhofer2020,Tao2025}, to  nuclear spins  \cite{Krithika2022} and trapped ions \cite{Zhang2023,Li2025,Liuvdp2025}.

A first observation of quantum phase locking to an external signal was made in  spin-1 systems in 2020 \cite{Laskar2020,Koppenhofer2020}. In one experiment, the system was realized  with three hyperfine states of laser-cooled Rb atoms with no initial phase relation \cite{Laskar2020}. 
  Laser fields were used to effectively implement incoherent loss and gain processes, producing a stable  limit-cycle state and quantum  synchronization of the phase difference of spin coherences. 
The experimentally determined Husimi $Q$ function \eqref{7}  showed entrainment to the drive (Fig.~\ref{fig:Laskar}), as well as a blockade of synchronization for balanced gain and loss, as predicted \cite{Roulet2018}. Similar results were obtained in another experimental simulation of a spin-1 system encoded in two superconducting qubits on the IBM-Q system \cite{Koppenhofer2020}. Robust quantum  entrainment was subsequently observed in a pair of interacting nuclear spins subjected to an external drive in a nuclear magnetic resonance architecture  \cite{Krithika2022}, as well as in  a single trapped-ion qubit \cite{Zhang2023}.

\begin{figure}[t]
  \centering
  \begin{tikzpicture}
     \node (a) at (0,0) {\includegraphics[scale=1.1]{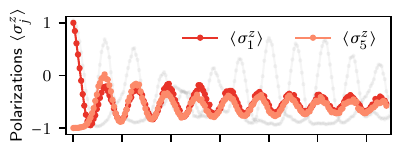}};
     \node (b) at (0,-2.9) {\includegraphics[scale=1.1]{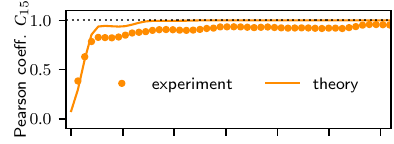}};
     \node (c) at (0,-6) {\includegraphics[scale=1.1]{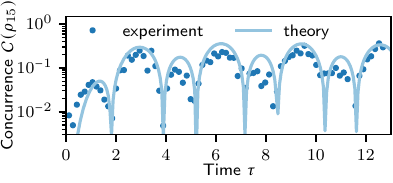}};
    \draw (-4,1) node {(a)};
  \draw (-4,-1.9) node {(b)};
   \draw (-4,-4.6) node {(c)};
  \end{tikzpicture}
  \caption{{Entangled noise-induced synchronization of superconducting qubits. (a)  Average dynamics of the local $z$-polarizations of the edge spins of an $N = 5$ spin chain, with Gaussian noise applied to the middle. (b) Corresponding Pearson correlation coefficient, \cref{eq:Pearson}, displaying the transition to synchrony. (c) Starting from a separable state, constant amplitude entanglement is generated between the end spins, as quantified by the concurrence of the tomographically reconstructed state (taken from Ref.~\cite{Tao2025}).}}
  \label{fig:Tao}
\end{figure}
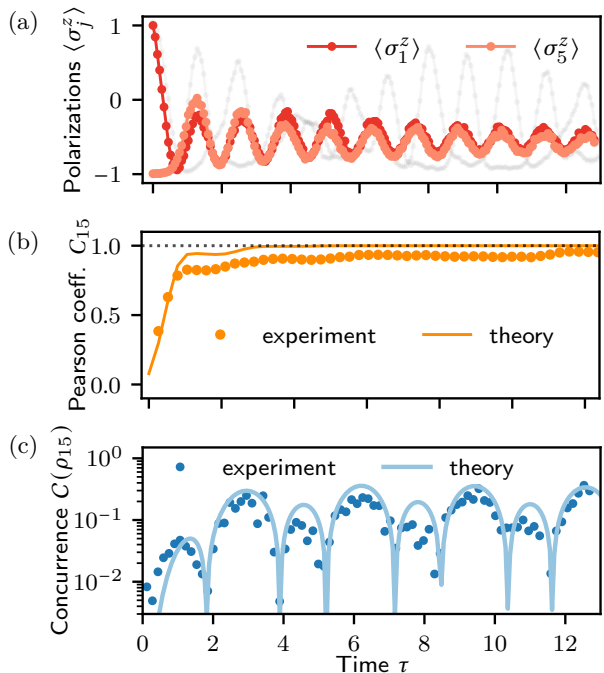

The first  realization of a quantum van der Pol oscillator, as theoretically described by Eq.~\eqref{eq:quantum-vdP}, was achieved in Ref.~\cite{Li2025} using a single trapped  ion.
The quantum oscillator was successfully phase synchronized to an external laser drive, and the enhancing effects of both squeezing and strong dissipation were explicitly demonstrated by means of the reconstructed Wigner function \eqref{6} (\cref{fig:Li}) {in line with theoretical predictions \cite{Sonar2018,Mok2020}}.  The bifurcation to a bistable phase-space distribution for large
squeezing was further demonstrated. The mutual synchronization  of two trapped-ion van der Pol oscillators, as well as their entrainment to an external drive,  were additionally examined in Ref.~\cite{Liuvdp2025}.

On the other hand, dynamical noise-induced quantum synchronization was observed  in a chain of superconducting transmon qubits with nearest-neighbor interactions and subject to Gaussian white noise applied at a single site \cite{Tao2025}. Entangled synchronized oscillations of edge spins were established by evaluating the Pearson correlation coefficient \eqref{eq:Pearson} of their averaged magnetization together with their concurrence (Fig.~\ref{fig:Tao}).
This experiment has initiated the exploration of collective synchronization effects with stronger than classical correlations.

\section{Many-body synchronization on quantum networks}

While experiments have so far focused on quantum synchronization in few-body systems, theory has recently explored many-body synchrony in quantum networks and the influence of their complex topologies.
In systems comprised of many quantum limit cycle oscillators, mean-field methods often result in Kuramoto-like effective descriptions of the semiclassical evolution of the individual phase variables that are modified by quantum effects \cite{Heinrich2011,Lee2013,Ludwig2013,Lee2014,Witthaut2017,Nadolny2023,Dai2026}. These systems feature a sharp transition from an incoherent to a fully or partially phase-coherent global state.
Such collective oscillator dynamics have, for instance, been studied in arrays of optomechanical cells that interact harmonically by coupling the mechanical degree of freedom \cite{Heinrich2011,Ludwig2013}.
The interaction between the laser-driven optical mode and the mechanical mode stabilizes self-sustained oscillations beyond a coupling threshold, and {their collective synchronization} can be witnessed by a localized Wigner function.
Whereas the semiclassical model features an antisymmetric Arnold-tongue, predicting both in-phase and anti-phase synchronized steady states, the interplay between optical and mechanical modes in the presence of cavity shot noise and Heisenberg uncertainty gives rise to a nonsymmetric, rich phase diagram as a function of detuning and coupling strength \cite{Heinrich2011,Ludwig2013}.

For a class of isolated many-body quantum systems with  on-site two-body interactions, synchronization of the mean-field Kuramoto description can even serve as an indicator for persistent entanglement in the quantum mechanical model \cite{Witthaut2017}. Moreover, synchronization of two groups of all-to-all coupled spin-1 van der Pol oscillator networks has been investigated in  Ref.~\cite{Nadolny2023}, where an exact mean-field treatment results in nonlinear Lindblad evolution in the thermodynamic limit.
It was found that synchronization blockades, due to interference and quantization, are inherited from the microscopic model, while additional phase frustration blockades exist that can be attributed only to macroscopic effects.

On the other hand, dynamical synchronization has been studied in a  network of harmonic {oscillators} with random  frequencies subject to common global dissipation by a thermal environment \cite{Manzano2013}.
By changing the frequency of a single node in this system, it is possible to create a decoherence-free subspace that is immune to the dissipation, and leads to coherent, synchronized oscillations of the entire network \cite{Manzano2013}.
In the subspace of only a single excitation, the topology of the underlying network completely determines the noiseless clusters \cite{Cabot2018}.
In this context, a major challenge is to find the right dissipation that leaves only the desired synchronizing modes invariant, while suppressing all other modes.
Results thus far suggest that the interplay between the topology of the physical network and the Hilbert space topology shape the synchronization properties of the system \cite{Manzano2013,Cabot2018,Tindall2020,Buca2022,Schmolke2022,Tong2025,Cakmak2026}.
In certain cases, for instance in the subspace of a single excitation, a one-to-one correspondence between the two can be established, where the synchronization conditions are determined by the adjacency matrix of the underlying graph \cite{Manzano2013,Cabot2018,Schmolke2022,Tong2025,Cakmak2026}.

Furthermore, the amount of decoherence-free eigenmodes that contribute to the synchronous dynamics is typically a small, finite number.
The magnitude of the synchronized oscillations therefore scales unfavorably with the system size, and may vanish in the thermodynamic limit \cite{Schmolke2022,Schmolke2024}.
Robustness against vanishing amplitudes and static disorder can be achieved by harnessing the Hilbert space topology.
By exploiting the topological edge states, stable synchronization of local observables has been demonstrated in a spin-$1$ Affleck--Kennedy--Lieb--Tasaki chain and in a Aubry--Andre--Harper model \cite{Wachtler2024,Tong2025}.
The topological order additionally  provides protection  against the breaking of the permutation symmetry of the physical network.

\section{Concluding perspective}

Classical synchronization theory has provided profound insights into collective behavior \cite{Kuramoto1984,Blekman1988,Mosekilde2002,Pikovsky2003,Strogatz2003,Acebron2005,Anishchenko2007,osi07,Balanov2009,Boccaletti2018}. Beyond explaining a wide range of natural phenomena, such as the coordinated activity of neurons in the brain, it plays a central role in engineering applications, including the stabilization of electric power grids and communication networks. Synchronization is also fundamental to computer science, where it enables the reliable operation of multiprocessor systems and distributed computing architectures.  More broadly, synchronization theory offers powerful tools for understanding the emergence of self-organized dynamics in classical complex systems. As a result, it has had a unique impact across both classical science and technology \cite{Kuramoto1984,Blekman1988,Mosekilde2002,Pikovsky2003,Strogatz2003,Acebron2005,Anishchenko2007,osi07,Balanov2009,Boccaletti2018}.
The emerging field of quantum synchronization has demonstrated that these concepts can be successfully extended to the quantum regime, revealing genuinely nonclassical forms of synchrony, including coherence-enhanced synchronization between far-detuned oscillators and synchronous oscillations exhibiting stronger-than-classical correlations. The latter effects suggest that quantum synchronization is able to  go beyond what is classically possible. Recent studies have also highlighted potential applications in synchronization-based quantum communication \cite{li17,plj17}, complex quantum networks \cite{Lohe2010,Nokkala2024}, and quantum metrology \cite{vai25}. Quantum synchrony therefore offers a promising avenue for exploring collective quantum phenomena and advancing synchronization-based quantum technologies \cite{Nielsen2010}, where long-lived coherence and quantum correlations constitute essential resources.

\textbf{Acknowledgements.} The authors acknowledge financial support from the Vector Foundation and the German Research Foundation DFG (FOR 2724).

\bibliographystyle{naturemag}
\bibliography{review}

\begin{thebibliography}{100}
\expandafter\ifx\csname url\endcsname\relax
  \def\url#1{\texttt{#1}}\fi
\expandafter\ifx\csname urlprefix\endcsname\relax\def\urlprefix{URL }\fi
\providecommand{\bibinfo}[2]{#2}
\providecommand{\eprint}[2][]{\url{#2}}

\bibitem{Kuramoto1984}
\bibinfo{author}{Kuramoto, Y.}
\newblock \emph{\bibinfo{title}{Chemical Oscillations, Waves, and Turbulence}} (\bibinfo{publisher}{Springer, Berlin}, \bibinfo{year}{1984}).

\bibitem{Blekman1988}
\bibinfo{author}{Blekman, I.~I.}
\newblock \emph{\bibinfo{title}{Synchronization in Science and Technology}} (\bibinfo{publisher}{ASME Press}, \bibinfo{address}{New York}, \bibinfo{year}{1988}).

\bibitem{Mosekilde2002}
\bibinfo{author}{Mosekilde, E.}, \bibinfo{author}{Maistrenko, Y.} \& \bibinfo{author}{Postnov, D.}
\newblock \emph{\bibinfo{title}{Chaotic Synchronization}} (\bibinfo{publisher}{World Scientific}, \bibinfo{address}{Singapore}, \bibinfo{year}{2002}).

\bibitem{Pikovsky2003}
\bibinfo{author}{Pikovsky, A.}, \bibinfo{author}{Rosenblum, M.} \& \bibinfo{author}{Kurths, J.}
\newblock \emph{\bibinfo{title}{Synchronization: A Universal Concept in Nonlinear Sciences}} (\bibinfo{publisher}{Cambridge University Press}, \bibinfo{address}{Cambridge}, \bibinfo{year}{2003}).

\bibitem{Strogatz2003}
\bibinfo{author}{Strogatz, S.~H.}
\newblock \emph{\bibinfo{title}{Sync: How Order Emerges from Chaos in the Universe, Nature, and Daily Life}} (\bibinfo{publisher}{Hyperion}, \bibinfo{address}{New York}, \bibinfo{year}{2003}).

\bibitem{Acebron2005}
\bibinfo{author}{Acebr{\'o}n, J.~A.}, \bibinfo{author}{Bonilla, L.~L.}, \bibinfo{author}{Vicente, C. J.~P.}, \bibinfo{author}{Ritort, F.} \& \bibinfo{author}{Spigler, R.}
\newblock \bibinfo{title}{The {K}uramoto model: {A} simple paradigm for synchronization phenomena}.
\newblock \emph{\bibinfo{journal}{Rev. Mod. Phys.}} \textbf{\bibinfo{volume}{77}}, \bibinfo{pages}{137--185} (\bibinfo{year}{2005}).

\bibitem{Anishchenko2007}
\bibinfo{author}{Anishchenko, V.~S.}, \bibinfo{author}{Astakhov, V.}, \bibinfo{author}{Vadivasova, T.}, \bibinfo{author}{Neiman, A.} \& \bibinfo{author}{Schimansky-Geier, L.}
\newblock \emph{\bibinfo{title}{Nonlinear Dynamics of Chaotic and Stochastic Systems}} (\bibinfo{publisher}{Springer}, \bibinfo{address}{Berlin}, \bibinfo{year}{2007}).

\bibitem{osi07}
\bibinfo{author}{Osipov, G.~V.}, \bibinfo{author}{Kurths, J.} \& \bibinfo{author}{Zhou, C.}
\newblock \emph{\bibinfo{title}{Synchronization in Oscillatory Networks}} (\bibinfo{publisher}{Springer, Berlin}, \bibinfo{year}{2007}).

\bibitem{Balanov2009}
\bibinfo{author}{Balanov, A.}, \bibinfo{author}{Janson, N.}, \bibinfo{author}{Postnov, D.} \& \bibinfo{author}{Sosnovtseva, O.}
\newblock \emph{\bibinfo{title}{Synchronization: From Simple to Complex}} (\bibinfo{publisher}{Springer}, \bibinfo{address}{Berlin}, \bibinfo{year}{2009}).

\bibitem{Boccaletti2018}
\bibinfo{author}{Boccaletti, S.}, \bibinfo{author}{Pisarchik, A.~N.}, \bibinfo{author}{del Genio, C.~I.} \& \bibinfo{author}{Amann, A.}
\newblock \emph{\bibinfo{title}{Synchronization. From Coupled Systems to Complex Networks}} (\bibinfo{publisher}{Cambridge University Press}, \bibinfo{address}{Cambridge}, \bibinfo{year}{2018}).

\bibitem{ser09}
\bibinfo{author}{Serpedin, E.} \& \bibinfo{author}{Chaudhari, O.~M.}
\newblock \emph{\bibinfo{title}{Synchronization Wireless Sensor Networks}} (\bibinfo{publisher}{Cambridge University Press}, \bibinfo{address}{Cambridge}, \bibinfo{year}{2009}).

\bibitem{lin17}
\bibinfo{author}{Ling, F.}
\newblock \emph{\bibinfo{title}{Synchronization in Digital Communication Systems}} (\bibinfo{publisher}{Cambridge University Press}, \bibinfo{address}{Cambridge}, \bibinfo{year}{2017}).

\bibitem{Choi2017}
\bibinfo{author}{Choi, H.-H.} \& \bibinfo{author}{Lee, J.-R.}
\newblock \bibinfo{title}{Principles, applications, and challenges of synchronization in nature for future mobile communication systems}.
\newblock \emph{\bibinfo{journal}{Mobile Inf. Syst.}} \textbf{\bibinfo{volume}{2017}}, \bibinfo{pages}{8932631} (\bibinfo{year}{2017}).

\bibitem{liu97}
\bibinfo{author}{Liu, C.}, \bibinfo{author}{Weaver, D.~R.}, \bibinfo{author}{Strogatz, S.~H.} \& \bibinfo{author}{Reppert, S.~M.}
\newblock \bibinfo{title}{Cellular construction of a circadian clock: period determination in the suprachiasmatic nuclei}.
\newblock \emph{\bibinfo{journal}{Cell}} \textbf{\bibinfo{volume}{91}}, \bibinfo{pages}{855--860} (\bibinfo{year}{1997}).

\bibitem{pan02}
\bibinfo{author}{Pantaleone, J.}
\newblock \bibinfo{title}{Synchronization of metronomes}.
\newblock \emph{\bibinfo{journal}{Am. J. Phys.}} \textbf{\bibinfo{volume}{70}}, \bibinfo{pages}{992--1000} (\bibinfo{year}{2002}).

\bibitem{nei02}
\bibinfo{author}{Neiman, A.~B.} \& \bibinfo{author}{Russell, D.~F.}
\newblock \bibinfo{title}{Synchronization of noise-induced bursts in noncoupled sensory neurons}.
\newblock \emph{\bibinfo{journal}{Phys. Rev. Lett.}} \textbf{\bibinfo{volume}{88}}, \bibinfo{pages}{138103} (\bibinfo{year}{2002}).

\bibitem{poo03}
\bibinfo{author}{Poole, C.~P.} \& \bibinfo{author}{Owens, F.~J.}
\newblock \emph{\bibinfo{title}{Introduction to Nanotechnology}} (\bibinfo{publisher}{Wiley}, \bibinfo{address}{New York}, \bibinfo{year}{2003}).

\bibitem{Nielsen2010}
\bibinfo{author}{Nielsen, M.~A.} \& \bibinfo{author}{Chuang, I.~L.}
\newblock \emph{\bibinfo{title}{Quantum computation and quantum information}} (\bibinfo{publisher}{Cambridge University Press, Cambridge}, \bibinfo{year}{2010}).

\bibitem{Goychuk2006}
\bibinfo{author}{Goychuk, I.}, \bibinfo{author}{Casado-Pascual, J.}, \bibinfo{author}{Morillo, M.}, \bibinfo{author}{Lehmann, J.} \& \bibinfo{author}{H\"anggi, P.}
\newblock \bibinfo{title}{Quantum stochastic synchronization}.
\newblock \emph{\bibinfo{journal}{Phys. Rev. Lett.}} \textbf{\bibinfo{volume}{97}}, \bibinfo{pages}{210601} (\bibinfo{year}{2006}).

\bibitem{Zhirov2006}
\bibinfo{author}{Zhirov, O.~V.} \& \bibinfo{author}{Shepelyansky, D.~L.}
\newblock \bibinfo{title}{Quantum synchronization}.
\newblock \emph{\bibinfo{journal}{Eur. Phys. J. D}} \textbf{\bibinfo{volume}{38}}, \bibinfo{pages}{375--379} (\bibinfo{year}{2006}).

\bibitem{Zhirov2008}
\bibinfo{author}{Zhirov, O.~V.} \& \bibinfo{author}{Shepelyansky, D.~L.}
\newblock \bibinfo{title}{Synchronization and bistability of a qubit coupled to a driven dissipative oscillator}.
\newblock \emph{\bibinfo{journal}{Phys. Rev. Lett.}} \textbf{\bibinfo{volume}{100}}, \bibinfo{pages}{014101} (\bibinfo{year}{2008}).

\bibitem{arn14}
\bibinfo{author}{Arndt, M.} \& \bibinfo{author}{Hornberger, K.}
\newblock \bibinfo{title}{Testing the limits of quantum mechanical superpositions}.
\newblock \emph{\bibinfo{journal}{Nature Phys.}} \textbf{\bibinfo{volume}{10}}, \bibinfo{pages}{271--277} (\bibinfo{year}{2014}).

\bibitem{str17}
\bibinfo{author}{Streltsov, A.}, \bibinfo{author}{Adesso, G.} \& \bibinfo{author}{Plenio, M.~B.}
\newblock \bibinfo{title}{Colloquium: Quantum coherence as a resource}.
\newblock \emph{\bibinfo{journal}{Rev. Mod. Phys.}} \textbf{\bibinfo{volume}{89}}, \bibinfo{pages}{041003} (\bibinfo{year}{2017}).

\bibitem{hor09}
\bibinfo{author}{Horodecki, R.}, \bibinfo{author}{Horodecki, P.}, \bibinfo{author}{Horodecki, M.} \& \bibinfo{author}{Horodecki, K.}
\newblock \bibinfo{title}{Quantum entanglement}.
\newblock \emph{\bibinfo{journal}{Rev. Mod. Phys.}} \textbf{\bibinfo{volume}{81}}, \bibinfo{pages}{865--942} (\bibinfo{year}{2009}).

\bibitem{mod12}
\bibinfo{author}{Modi, K.}, \bibinfo{author}{Brodutch, A.}, \bibinfo{author}{Cable, H.}, \bibinfo{author}{Paterek, T.} \& \bibinfo{author}{Vedral, V.}
\newblock \bibinfo{title}{The classical-quantum boundary for correlations: discord and related measures}.
\newblock \emph{\bibinfo{journal}{Rev. Mod. Phys.}} \textbf{\bibinfo{volume}{84}}, \bibinfo{pages}{1655--1707} (\bibinfo{year}{2012}).

\bibitem{hil84}
\bibinfo{author}{Hillery, M. O. S.~M.}, \bibinfo{author}{O'Connell, R.~F.}, \bibinfo{author}{Scully, M.~O.} \& \bibinfo{author}{Wigner, E.~P.}
\newblock \bibinfo{title}{Distribution functions in physics: Fundamentals}.
\newblock \emph{\bibinfo{journal}{Phys. Rep.}} \textbf{\bibinfo{volume}{106}}, \bibinfo{pages}{121--167} (\bibinfo{year}{1984}).

\bibitem{Strogatz2024}
\bibinfo{author}{Strogatz, S.~H.}
\newblock \emph{\bibinfo{title}{Nonlinear Dynamics and Chaos}} (\bibinfo{publisher}{CRC Press}, \bibinfo{address}{Boca Raton}, \bibinfo{year}{2024}).

\bibitem{lyn95}
\bibinfo{author}{Lynch, R.}
\newblock \bibinfo{title}{The quantum phase problem: a critical review}.
\newblock \emph{\bibinfo{journal}{Phys. Rep.}} \textbf{\bibinfo{volume}{256}}, \bibinfo{pages}{367--436} (\bibinfo{year}{1995}).

\bibitem{Lee2013}
\bibinfo{author}{Lee, T.~E.} \& \bibinfo{author}{Sadeghpour, H.~R.}
\newblock \bibinfo{title}{Quantum synchronization of quantum van der {P}ol oscillators with trapped ions}.
\newblock \emph{\bibinfo{journal}{Phys. Rev. Lett.}} \textbf{\bibinfo{volume}{111}}, \bibinfo{pages}{234101} (\bibinfo{year}{2013}).

\bibitem{Walter2014}
\bibinfo{author}{Walter, S.}, \bibinfo{author}{Nunnenkamp, A.} \& \bibinfo{author}{Bruder, C.}
\newblock \bibinfo{title}{Quantum synchronization of a driven self-sustained oscillator}.
\newblock \emph{\bibinfo{journal}{Phys. Rev. Lett.}} \textbf{\bibinfo{volume}{112}}, \bibinfo{pages}{094102} (\bibinfo{year}{2014}).

\bibitem{jen13}
\bibinfo{author}{Jenkins, A.}
\newblock \bibinfo{title}{Self-oscillation}.
\newblock \emph{\bibinfo{journal}{Phys. Rep.}} \textbf{\bibinfo{volume}{525}}, \bibinfo{pages}{167--222} (\bibinfo{year}{2013}).

\bibitem{Rayleigh1883}
\bibinfo{author}{Rayleigh, L.}
\newblock \bibinfo{title}{{XXXIII}. {O}n maintained vibrations}.
\newblock \emph{\bibinfo{journal}{London Edinburgh Dublin Philos. Mag. \& J. Sci.}} \textbf{\bibinfo{volume}{15}}, \bibinfo{pages}{229--235} (\bibinfo{year}{1883}).

\bibitem{vanderPol1927}
\bibinfo{author}{van~der Pol~Jun., B.}
\newblock \bibinfo{title}{{VII}. {F}orced oscillations in a circuit with non-linear resistance. (reception with reactive triode)}.
\newblock \emph{\bibinfo{journal}{London Edinburgh Dublin Philos. Mag. \& J. Sci.}} \textbf{\bibinfo{volume}{3}}, \bibinfo{pages}{65--80} (\bibinfo{year}{1927}).

\bibitem{Lee2014}
\bibinfo{author}{Lee, T.~E.}, \bibinfo{author}{Chan, C.-K.} \& \bibinfo{author}{Wang, S.}
\newblock \bibinfo{title}{Entanglement tongue and quantum synchronization of disordered oscillators}.
\newblock \emph{\bibinfo{journal}{Phys. Rev. E}} \textbf{\bibinfo{volume}{89}}, \bibinfo{pages}{022913} (\bibinfo{year}{2014}).

\bibitem{Lorch2016}
\bibinfo{author}{L\"orch, N.}, \bibinfo{author}{Amitai, E.}, \bibinfo{author}{Nunnenkamp, A.} \& \bibinfo{author}{Bruder, C.}
\newblock \bibinfo{title}{Genuine quantum signatures in synchronization of anharmonic self-oscillators}.
\newblock \emph{\bibinfo{journal}{Phys. Rev. Lett.}} \textbf{\bibinfo{volume}{117}}, \bibinfo{pages}{073601} (\bibinfo{year}{2016}).

\bibitem{Lorch2017}
\bibinfo{author}{L\"orch, N.}, \bibinfo{author}{Nigg, S.~E.}, \bibinfo{author}{Nunnenkamp, A.}, \bibinfo{author}{Tiwari, R.~P.} \& \bibinfo{author}{Bruder, C.}
\newblock \bibinfo{title}{Quantum synchronization blockade: Energy quantization hinders synchronization of identical oscillators}.
\newblock \emph{\bibinfo{journal}{Phys. Rev. Lett.}} \textbf{\bibinfo{volume}{118}}, \bibinfo{pages}{243602} (\bibinfo{year}{2017}).

\bibitem{Sonar2018}
\bibinfo{author}{Sonar, S.} \emph{et~al.}
\newblock \bibinfo{title}{Squeezing enhances quantum synchronization}.
\newblock \emph{\bibinfo{journal}{Phys. Rev. Lett.}} \textbf{\bibinfo{volume}{120}}, \bibinfo{pages}{163601} (\bibinfo{year}{2018}).

\bibitem{Mok2020}
\bibinfo{author}{Mok, W.-K.}, \bibinfo{author}{Kwek, L.-C.} \& \bibinfo{author}{Heimonen, H.}
\newblock \bibinfo{title}{Synchronization boost with single-photon dissipation in the deep quantum regime}.
\newblock \emph{\bibinfo{journal}{Phys. Rev. Res.}} \textbf{\bibinfo{volume}{2}}, \bibinfo{pages}{033422} (\bibinfo{year}{2020}).

\bibitem{Arosh2021}
\bibinfo{author}{Ben~Arosh, L.}, \bibinfo{author}{Cross, M.~C.} \& \bibinfo{author}{Lifshitz, R.}
\newblock \bibinfo{title}{Quantum limit cycles and the {R}ayleigh and van der {P}ol oscillators}.
\newblock \emph{\bibinfo{journal}{Phys. Rev. Res.}} \textbf{\bibinfo{volume}{3}}, \bibinfo{pages}{013130} (\bibinfo{year}{2021}).

\bibitem{Nadolny2023}
\bibinfo{author}{Nadolny, T.} \& \bibinfo{author}{Bruder, C.}
\newblock \bibinfo{title}{Macroscopic quantum synchronization effects}.
\newblock \emph{\bibinfo{journal}{Phys. Rev. Lett.}} \textbf{\bibinfo{volume}{131}}, \bibinfo{pages}{190402} (\bibinfo{year}{2023}).

\bibitem{Chia2020}
\bibinfo{author}{Chia, A.}, \bibinfo{author}{Kwek, L.~C.} \& \bibinfo{author}{Noh, C.}
\newblock \bibinfo{title}{Relaxation oscillations and frequency entrainment in quantum mechanics}.
\newblock \emph{\bibinfo{journal}{Phys. Rev. E}} \textbf{\bibinfo{volume}{102}}, \bibinfo{pages}{042213} (\bibinfo{year}{2020}).

\bibitem{Chia2025}
\bibinfo{author}{Chia, A.}, \bibinfo{author}{Mok, W.-K.}, \bibinfo{author}{Kwek, L.-C.} \& \bibinfo{author}{Noh, C.}
\newblock \bibinfo{title}{Quantization of nonlinear non-{H}amiltonian systems}.
\newblock \emph{\bibinfo{journal}{Phys. Rev. E}} \textbf{\bibinfo{volume}{112}}, \bibinfo{pages}{054206} (\bibinfo{year}{2025}).

\bibitem{Kehrer2025}
\bibinfo{author}{Kehrer, T.}, \bibinfo{author}{Bruder, C.} \& \bibinfo{author}{Solanki, P.}
\newblock \bibinfo{title}{Quantum synchronization of twin limit-cycle oscillators}.
\newblock \emph{\bibinfo{journal}{Phys. Rev. Lett.}} \textbf{\bibinfo{volume}{135}}, \bibinfo{pages}{063601} (\bibinfo{year}{2025}).

\bibitem{Breuer2007}
\bibinfo{author}{Breuer, H.-P.} \& \bibinfo{author}{Petruccione, F.}
\newblock \emph{\bibinfo{title}{The Theory of Open Quantum Systems}} (\bibinfo{publisher}{Oxford University Press, Oxford}, \bibinfo{year}{2007}).

\bibitem{Barchielli2009}
\bibinfo{author}{Barchielli, A.} \& \bibinfo{author}{Gregoratti, M.}
\newblock \emph{\bibinfo{title}{Quantum Trajectories and Measurements in Continuous Time}} (\bibinfo{publisher}{Springer, Berlin}, \bibinfo{year}{2009}).

\bibitem{Wiseman2009}
\bibinfo{author}{Wiseman, H.~M.} \& \bibinfo{author}{Milburn, G.~J.}
\newblock \emph{\bibinfo{title}{Quantum Measurement and Control}} (\bibinfo{publisher}{Cambridge University Press, Cambridge}, \bibinfo{year}{2009}).

\bibitem{Jacobs2014}
\bibinfo{author}{Jacobs, K.}
\newblock \emph{\bibinfo{title}{Quantum Measurement Theory and its Applications}} (\bibinfo{publisher}{Cambridge University Press, Cambridge}, \bibinfo{year}{2014}).

\bibitem{Jordan2024}
\bibinfo{author}{Jordan, A.~N.} \& \bibinfo{author}{Siddiqi, I.~A.}
\newblock \emph{\bibinfo{title}{Quantum Measurement: Theory and Practice}} (\bibinfo{publisher}{Cambridge University Press, Cambridge}, \bibinfo{year}{2024}).

\bibitem{Setoyama2024}
\bibinfo{author}{Setoyama, W.} \& \bibinfo{author}{Hasegawa, Y.}
\newblock \bibinfo{title}{Lie algebraic quantum phase reduction}.
\newblock \emph{\bibinfo{journal}{Phys. Rev. Lett.}} \textbf{\bibinfo{volume}{132}}, \bibinfo{pages}{093602} (\bibinfo{year}{2024}).

\bibitem{Setoyama2025}
\bibinfo{author}{Setoyama, W.} \& \bibinfo{author}{Hasegawa, Y.}
\newblock \bibinfo{title}{Lie-algebraic quantum phase reduction based on heterodyne detection}.
\newblock \emph{\bibinfo{journal}{Phys. Rev. A}} \textbf{\bibinfo{volume}{111}}, \bibinfo{pages}{012202} (\bibinfo{year}{2025}).

\bibitem{YiZhao2025}
\bibinfo{author}{Zhao, Y.~J.}, \bibinfo{author}{Moore, J.~E.}, \bibinfo{author}{Thingna, J.} \& \bibinfo{author}{W\"achtler, C.~W.}
\newblock \bibinfo{title}{Quantum synchronization of perturbed oscillating coherences}.
\newblock \emph{\bibinfo{journal}{arXiv:2510.11601}}  (\bibinfo{year}{2025}).

\bibitem{Nadolny2026}
\bibinfo{author}{Nadolny, T.} \& \bibinfo{author}{Bruder, C.}
\newblock \bibinfo{title}{Quantum limit cycles and synchronization from a measurement perspective}.
\newblock \emph{\bibinfo{journal}{Phys. Rev. Res.}} \textbf{\bibinfo{volume}{8}}, \bibinfo{pages}{023050} (\bibinfo{year}{2026}).

\bibitem{Chen2026}
\bibinfo{author}{Chen, S.} \& \bibinfo{author}{Clerk, A.~A.}
\newblock \bibinfo{title}{Quantum limit cycles with continuous symmetries from coherent parametric driving: exact solutions and many-body extensions}.
\newblock \emph{\bibinfo{journal}{arXiv:2604.25864}}  (\bibinfo{year}{2026}).

\bibitem{Hassler2026}
\bibinfo{author}{Hassler, F.}, \bibinfo{author}{Scheer, D.}, \bibinfo{author}{Saquaque, S.} \& \bibinfo{author}{Kim, S.}
\newblock \bibinfo{title}{Quantum synchronization of fock states}.
\newblock \emph{\bibinfo{journal}{arXiv:2605.30271}}  (\bibinfo{year}{2026}).

\bibitem{Christiansen2026}
\bibinfo{author}{Christiansen, H.} \& \bibinfo{author}{Paaske, J.}
\newblock \bibinfo{title}{Quantum desynchronization of limit cycles}.
\newblock \emph{\bibinfo{journal}{arXiv:2605.30302}}  (\bibinfo{year}{2026}).

\bibitem{hus15}
\bibinfo{author}{Hush, M.~R.}, \bibinfo{author}{Li, W.}, \bibinfo{author}{Genway, S.}, \bibinfo{author}{Lesanovsky, I.} \& \bibinfo{author}{Armour, A.~D.}
\newblock \bibinfo{title}{Spin correlations as a probe of quantum synchronization in trapped-ion phonon lasers}.
\newblock \emph{\bibinfo{journal}{Phys. Rev. A}} \textbf{\bibinfo{volume}{91}}, \bibinfo{pages}{061401(R)} (\bibinfo{year}{2015}).

\bibitem{Heinrich2011}
\bibinfo{author}{Heinrich, G.}, \bibinfo{author}{Ludwig, M.}, \bibinfo{author}{Qian, J.}, \bibinfo{author}{Kubala, B.} \& \bibinfo{author}{Marquardt, F.}
\newblock \bibinfo{title}{Collective dynamics in optomechanical arrays}.
\newblock \emph{\bibinfo{journal}{Phys. Rev. Lett.}} \textbf{\bibinfo{volume}{107}}, \bibinfo{pages}{043603} (\bibinfo{year}{2011}).

\bibitem{Ludwig2013}
\bibinfo{author}{Ludwig, M.} \& \bibinfo{author}{Marquardt, F.}
\newblock \bibinfo{title}{Quantum many-body dynamics in optomechanical arrays}.
\newblock \emph{\bibinfo{journal}{Phys. Rev. Lett.}} \textbf{\bibinfo{volume}{111}}, \bibinfo{pages}{073603} (\bibinfo{year}{2013}).

\bibitem{Weiss2016}
\bibinfo{author}{Weiss, T.}, \bibinfo{author}{Kronwald, A.} \& \bibinfo{author}{Marquardt, F.}
\newblock \bibinfo{title}{Noise-induced transitions in optomechanical synchronization}.
\newblock \emph{\bibinfo{journal}{New J. Phys.}} \textbf{\bibinfo{volume}{18}}, \bibinfo{pages}{013043} (\bibinfo{year}{2016}).

\bibitem{Roulet2018a}
\bibinfo{author}{Roulet, A.} \& \bibinfo{author}{Bruder, C.}
\newblock \bibinfo{title}{Synchronizing the smallest possible system}.
\newblock \emph{\bibinfo{journal}{Phys. Rev. Lett.}} \textbf{\bibinfo{volume}{121}}, \bibinfo{pages}{053601} (\bibinfo{year}{2018}).

\bibitem{Roulet2018}
\bibinfo{author}{Roulet, A.} \& \bibinfo{author}{Bruder, C.}
\newblock \bibinfo{title}{Quantum synchronization and entanglement generation}.
\newblock \emph{\bibinfo{journal}{Phys. Rev. Lett.}} \textbf{\bibinfo{volume}{121}}, \bibinfo{pages}{063601} (\bibinfo{year}{2018}).

\bibitem{Koppenhofer2019}
\bibinfo{author}{Koppenh\"ofer, M.} \& \bibinfo{author}{Roulet, A.}
\newblock \bibinfo{title}{Optimal synchronization deep in the quantum regime: Resource and fundamental limit}.
\newblock \emph{\bibinfo{journal}{Phys. Rev. A}} \textbf{\bibinfo{volume}{99}}, \bibinfo{pages}{043804} (\bibinfo{year}{2019}).

\bibitem{Lopez2020}
\bibinfo{author}{Parra-L\'opez, A.} \& \bibinfo{author}{Bergli, J.}
\newblock \bibinfo{title}{Synchronization in two-level quantum systems}.
\newblock \emph{\bibinfo{journal}{Phys. Rev. A}} \textbf{\bibinfo{volume}{101}}, \bibinfo{pages}{062104} (\bibinfo{year}{2020}).

\bibitem{Kehrer2024}
\bibinfo{author}{Kehrer, T.}, \bibinfo{author}{Nadolny, T.} \& \bibinfo{author}{Bruder, C.}
\newblock \bibinfo{title}{Quantum synchronization through the interference blockade}.
\newblock \emph{\bibinfo{journal}{Phys. Rev. A}} \textbf{\bibinfo{volume}{110}}, \bibinfo{pages}{042203} (\bibinfo{year}{2024}).

\bibitem{Kato2025}
\bibinfo{author}{Kato, Y.} \& \bibinfo{author}{Nakao, H.}
\newblock \bibinfo{title}{Quantum spin van der pol oscillator: Spin-based limit-cycle oscillator exhibiting quantum synchronization}.
\newblock \emph{\bibinfo{journal}{Phys. Rev. Res.}} \textbf{\bibinfo{volume}{7}}, \bibinfo{pages}{043133} (\bibinfo{year}{2025}).

\bibitem{Gilmore1975}
\bibinfo{author}{Gilmore, R.}, \bibinfo{author}{Bowden, C.~M.} \& \bibinfo{author}{Narducci, L.~M.}
\newblock \bibinfo{title}{Classical-quantum correspondence for multilevel systems}.
\newblock \emph{\bibinfo{journal}{Phys. Rev. A}} \textbf{\bibinfo{volume}{12}}, \bibinfo{pages}{1019--1031} (\bibinfo{year}{1975}).

\bibitem{Chryssomalakos2018}
\bibinfo{author}{Chryssomalakos, C.}, \bibinfo{author}{Guzman-Gonzalez, E.} \& \bibinfo{author}{Serrano-Ensastiga, E.}
\newblock \bibinfo{title}{Geometry of spin coherent states}.
\newblock \emph{\bibinfo{journal}{J. Phys. A: Math. Theor.}} \textbf{\bibinfo{volume}{51}}, \bibinfo{pages}{165202} (\bibinfo{year}{2018}).

\bibitem{Kam2023}
\bibinfo{author}{Kam, C.-F.}, \bibinfo{author}{Zhang, W.-M.} \& \bibinfo{author}{Feng, D.-H.}
\newblock \emph{\bibinfo{title}{Coherent States}} (\bibinfo{publisher}{Springer, Berlin}, \bibinfo{year}{2023}).

\bibitem{Nakao2016}
\bibinfo{author}{Nakao, H.}
\newblock \bibinfo{title}{Phase reduction approach to synchronisation of nonlinear oscillators}.
\newblock \emph{\bibinfo{journal}{Contemp. Phys.}} \textbf{\bibinfo{volume}{57}}, \bibinfo{pages}{188--214} (\bibinfo{year}{2016}).

\bibitem{Walter2015}
\bibinfo{author}{Walter, S.}, \bibinfo{author}{Nunnenkamp, A.} \& \bibinfo{author}{Bruder, C.}
\newblock \bibinfo{title}{Quantum synchronization of two van der {P}ol oscillators}.
\newblock \emph{\bibinfo{journal}{Ann. Phys.}} \textbf{\bibinfo{volume}{527}}, \bibinfo{pages}{131--138} (\bibinfo{year}{2015}).

\bibitem{Tilley2018}
\bibinfo{author}{Davis-Tilley, C.}, \bibinfo{author}{Teoh, C.~K.} \& \bibinfo{author}{Armour, A.~D.}
\newblock \bibinfo{title}{Dynamics of many-body quantum synchronisation}.
\newblock \emph{\bibinfo{journal}{New J. Phys.}} \textbf{\bibinfo{volume}{20}}, \bibinfo{pages}{113002} (\bibinfo{year}{2018}).

\bibitem{Orth2010}
\bibinfo{author}{Orth, P.~P.}, \bibinfo{author}{Roosen, D.}, \bibinfo{author}{Hofstetter, W.} \& \bibinfo{author}{Le~Hur, K.}
\newblock \bibinfo{title}{Dynamics, synchronization, and quantum phase transitions of two dissipative spins}.
\newblock \emph{\bibinfo{journal}{Phys. Rev. B}} \textbf{\bibinfo{volume}{82}}, \bibinfo{pages}{144423} (\bibinfo{year}{2010}).

\bibitem{Giorgi2012}
\bibinfo{author}{Giorgi, G.~L.}, \bibinfo{author}{Galve, F.}, \bibinfo{author}{Manzano, G.}, \bibinfo{author}{Colet, P.} \& \bibinfo{author}{Zambrini, R.}
\newblock \bibinfo{title}{Quantum correlations and mutual synchronization}.
\newblock \emph{\bibinfo{journal}{Phys. Rev. A}} \textbf{\bibinfo{volume}{85}}, \bibinfo{pages}{052101} (\bibinfo{year}{2012}).

\bibitem{Giorgi2013}
\bibinfo{author}{Giorgi, G.~L.}, \bibinfo{author}{Plastina, F.}, \bibinfo{author}{Francica, G.} \& \bibinfo{author}{Zambrini, R.}
\newblock \bibinfo{title}{Spontaneous synchronization and quantum correlation dynamics of open spin systems}.
\newblock \emph{\bibinfo{journal}{Phys. Rev. A}} \textbf{\bibinfo{volume}{88}}, \bibinfo{pages}{042115} (\bibinfo{year}{2013}).

\bibitem{Galve2017}
\bibinfo{author}{Galve, F.}, \bibinfo{author}{Luca~Giorgi, G.} \& \bibinfo{author}{Zambrini, R.}
\newblock \emph{\bibinfo{title}{Quantum Correlations and Synchronization Measures}}, \bibinfo{pages}{393--420} (\bibinfo{publisher}{Springer, Berlin}, \bibinfo{year}{2017}).

\bibitem{Bellomo2017}
\bibinfo{author}{Bellomo, B.}, \bibinfo{author}{Giorgi, G.~L.}, \bibinfo{author}{Palma, G.~M.} \& \bibinfo{author}{Zambrini, R.}
\newblock \bibinfo{title}{Quantum synchronization as a local signature of super- and subradiance}.
\newblock \emph{\bibinfo{journal}{Phys. Rev. A}} \textbf{\bibinfo{volume}{95}}, \bibinfo{pages}{043807} (\bibinfo{year}{2017}).

\bibitem{Karpat2019}
\bibinfo{author}{Karpat, G.}, \bibinfo{author}{Yal\c{c}\i~nkaya, I.} \& \bibinfo{author}{\c{C}akmak, B.}
\newblock \bibinfo{title}{Quantum synchronization in a collision model}.
\newblock \emph{\bibinfo{journal}{Phys. Rev. A}} \textbf{\bibinfo{volume}{100}}, \bibinfo{pages}{012133} (\bibinfo{year}{2019}).

\bibitem{Cabot2019}
\bibinfo{author}{Cabot, A.}, \bibinfo{author}{Giorgi, G.~L.}, \bibinfo{author}{Galve, F.} \& \bibinfo{author}{Zambrini, R.}
\newblock \bibinfo{title}{Quantum synchronization in dimer atomic lattices}.
\newblock \emph{\bibinfo{journal}{Phys. Rev. Lett.}} \textbf{\bibinfo{volume}{123}}, \bibinfo{pages}{023604} (\bibinfo{year}{2019}).

\bibitem{Giorgi2019}
\bibinfo{author}{Giorgi, G.~L.}, \bibinfo{author}{Cabot, A.} \& \bibinfo{author}{Zambrini, R.}
\newblock \emph{\bibinfo{title}{Transient Synchronization in Open Quantum Systems}}, \bibinfo{pages}{73--89} (\bibinfo{publisher}{Springer, Berlin}, \bibinfo{year}{2019}).

\bibitem{Buca2022}
\bibinfo{author}{Buca, B.}, \bibinfo{author}{Booker, C.} \& \bibinfo{author}{Jaksch, D.}
\newblock \bibinfo{title}{{Algebraic theory of quantum synchronization and limit cycles under dissipation}}.
\newblock \emph{\bibinfo{journal}{SciPost Phys.}} \textbf{\bibinfo{volume}{12}}, \bibinfo{pages}{097} (\bibinfo{year}{2022}).

\bibitem{Schmolke2022}
\bibinfo{author}{Schmolke, F.} \& \bibinfo{author}{Lutz, E.}
\newblock \bibinfo{title}{Noise-induced quantum synchronization}.
\newblock \emph{\bibinfo{journal}{Phys. Rev. Lett.}} \textbf{\bibinfo{volume}{129}}, \bibinfo{pages}{250601} (\bibinfo{year}{2022}).

\bibitem{Schmolke2024}
\bibinfo{author}{Schmolke, F.} \& \bibinfo{author}{Lutz, E.}
\newblock \bibinfo{title}{Measurement-induced quantum synchronization and multiplexing}.
\newblock \emph{\bibinfo{journal}{Phys. Rev. Lett.}} \textbf{\bibinfo{volume}{132}}, \bibinfo{pages}{010402} (\bibinfo{year}{2024}).

\bibitem{Barlow1993}
\bibinfo{author}{Barlow, R.~J.}
\newblock \emph{\bibinfo{title}{Statistics}} (\bibinfo{publisher}{Wiley, New York}, \bibinfo{year}{1993}).

\bibitem{Stefanini2025}
\bibinfo{author}{Stefanini, M.} \emph{et~al.}
\newblock \bibinfo{title}{Is {L}indblad for me?}
\newblock \emph{\bibinfo{journal}{arXiv:2506.22436}}  (\bibinfo{year}{2025}).

\bibitem{Gyamfi2020}
\bibinfo{author}{Gyamfi, J.~A.}
\newblock \bibinfo{title}{Fundamentals of quantum mechanics in liouville space}.
\newblock \emph{\bibinfo{journal}{Euro. J. Phys.}} \textbf{\bibinfo{volume}{41}}, \bibinfo{pages}{063002} (\bibinfo{year}{2020}).

\bibitem{Lidar1998}
\bibinfo{author}{Lidar, D.~A.}, \bibinfo{author}{Chuang, I.~L.} \& \bibinfo{author}{Whaley, K.~B.}
\newblock \bibinfo{title}{Decoherence-free subspaces for quantum computation}.
\newblock \emph{\bibinfo{journal}{Phys. Rev. Lett.}} \textbf{\bibinfo{volume}{81}}, \bibinfo{pages}{2594--2597} (\bibinfo{year}{1998}).

\bibitem{Knill2000}
\bibinfo{author}{Knill, E.}, \bibinfo{author}{Laflamme, R.} \& \bibinfo{author}{Viola, L.}
\newblock \bibinfo{title}{Theory of quantum error correction for general noise}.
\newblock \emph{\bibinfo{journal}{Phys. Rev. Lett.}} \textbf{\bibinfo{volume}{84}}, \bibinfo{pages}{2525--2528} (\bibinfo{year}{2000}).

\bibitem{Baumgartner2008}
\bibinfo{author}{Baumgartner, B.} \& \bibinfo{author}{Narnhofer, H.}
\newblock \bibinfo{title}{Analysis of quantum semigroups with {GKS}-{L}indblad generators: {II}. general}.
\newblock \emph{\bibinfo{journal}{J. Phys. A: Math. Theor.}} \textbf{\bibinfo{volume}{41}}, \bibinfo{pages}{395303} (\bibinfo{year}{2008}).

\bibitem{Baumgartner2012}
\bibinfo{author}{Baumgartner, B.} \& \bibinfo{author}{Narnhofer, H.}
\newblock \bibinfo{title}{The structures of state space concerning quantum dynamical semigroups}.
\newblock \emph{\bibinfo{journal}{Rev. Math. Phys.}} \textbf{\bibinfo{volume}{24}}, \bibinfo{pages}{1250001} (\bibinfo{year}{2012}).

\bibitem{Blume2010}
\bibinfo{author}{Blume-Kohout, R.}, \bibinfo{author}{Ng, H.~K.}, \bibinfo{author}{Poulin, D.} \& \bibinfo{author}{Viola, L.}
\newblock \bibinfo{title}{Information-preserving structures: A general framework for quantum zero-error information}.
\newblock \emph{\bibinfo{journal}{Phys. Rev. A}} \textbf{\bibinfo{volume}{82}}, \bibinfo{pages}{062306} (\bibinfo{year}{2010}).

\bibitem{Deutsch1991}
\bibinfo{author}{Deutsch, J.~M.}
\newblock \bibinfo{title}{Quantum statistical mechanics in a closed system}.
\newblock \emph{\bibinfo{journal}{Phys. Rev. A}} \textbf{\bibinfo{volume}{43}}, \bibinfo{pages}{2046--2049} (\bibinfo{year}{1991}).

\bibitem{Srednicki1999}
\bibinfo{author}{Srednicki, M.}
\newblock \bibinfo{title}{The approach to thermal equilibrium in quantized chaotic systems}.
\newblock \emph{\bibinfo{journal}{J. Phys. A: Math. Gen.}} \textbf{\bibinfo{volume}{32}}, \bibinfo{pages}{1163} (\bibinfo{year}{1999}).

\bibitem{DAlessio2016}
\bibinfo{author}{D'Alessio, L.}, \bibinfo{author}{Kafri, Y.}, \bibinfo{author}{Polkovnikov, A.} \& \bibinfo{author}{Rigol, M.}
\newblock \bibinfo{title}{From quantum chaos and eigenstate thermalization to statistical mechanics and thermodynamics}.
\newblock \emph{\bibinfo{journal}{Adv. Phys.}} \textbf{\bibinfo{volume}{65}}, \bibinfo{pages}{239--362} (\bibinfo{year}{2016}).

\bibitem{Deutsch2018}
\bibinfo{author}{Deutsch, J.~M.}
\newblock \bibinfo{title}{Eigenstate thermalization hypothesis}.
\newblock \emph{\bibinfo{journal}{Rep. Progr. Phys.}} \textbf{\bibinfo{volume}{81}}, \bibinfo{pages}{082001} (\bibinfo{year}{2018}).

\bibitem{Albert2018}
\bibinfo{author}{Albert, V.~V.}
\newblock \bibinfo{title}{Lindbladians with multiple steady states: theory and applications}.
\newblock \emph{\bibinfo{journal}{arXiv:1802.00010}}  (\bibinfo{year}{2018}).

\bibitem{Serbyn2021}
\bibinfo{author}{Serbyn, M.}, \bibinfo{author}{Abanin, D.~A.} \& \bibinfo{author}{Papic, Z.}
\newblock \bibinfo{title}{Quantum many-body scars and weak breaking of ergodicity}.
\newblock \emph{\bibinfo{journal}{Nature Phys.}} \textbf{\bibinfo{volume}{17}}, \bibinfo{pages}{675--685} (\bibinfo{year}{2021}).

\bibitem{Harrington2022}
\bibinfo{author}{Harrington, P.~M.}, \bibinfo{author}{Mueller, E.~J.} \& \bibinfo{author}{Murch, K.~W.}
\newblock \bibinfo{title}{Engineered dissipation for quantum information science}.
\newblock \emph{\bibinfo{journal}{Nature Reviews Physics}} \textbf{\bibinfo{volume}{4}}, \bibinfo{pages}{660–671} (\bibinfo{year}{2022}).

\bibitem{Karpat2020}
\bibinfo{author}{Karpat, G.}, \bibinfo{author}{Yal\c{c}{\i}nkaya, I.} \& \bibinfo{author}{\c{C}akmak, B.}
\newblock \bibinfo{title}{Quantum synchronization of few-body systems under collective dissipation}.
\newblock \emph{\bibinfo{journal}{Phys. Rev. A}} \textbf{\bibinfo{volume}{101}}, \bibinfo{pages}{042121} (\bibinfo{year}{2020}).

\bibitem{Goldobin2005}
\bibinfo{author}{Goldobin, D.} \& \bibinfo{author}{Pikovsky, A.}
\newblock \bibinfo{title}{Synchronization of self-sustained oscillators by common white noise}.
\newblock \emph{\bibinfo{journal}{Physica A}} \textbf{\bibinfo{volume}{351}}, \bibinfo{pages}{126--132} (\bibinfo{year}{2005}).

\bibitem{Nakao2007}
\bibinfo{author}{Nakao, H.}, \bibinfo{author}{Arai, K.} \& \bibinfo{author}{Kawamura, Y.}
\newblock \bibinfo{title}{Noise-induced synchronization and clustering in ensembles of uncoupled limit-cycle oscillators}.
\newblock \emph{\bibinfo{journal}{Phys. Rev. Lett.}} \textbf{\bibinfo{volume}{98}}, \bibinfo{pages}{184101} (\bibinfo{year}{2007}).

\bibitem{Teramae2008}
\bibinfo{author}{Teramae, J.-n.} \& \bibinfo{author}{Fukai, T.}
\newblock \bibinfo{title}{Temporal precision of spike response to fluctuating input in pulse-coupled networks of oscillating neurons}.
\newblock \emph{\bibinfo{journal}{Phys. Rev. Lett.}} \textbf{\bibinfo{volume}{101}}, \bibinfo{pages}{248105} (\bibinfo{year}{2008}).

\bibitem{Sunada2014}
\bibinfo{author}{Sunada, S.}, \bibinfo{author}{Arai, K.}, \bibinfo{author}{Yoshimura, K.} \& \bibinfo{author}{Adachi, M.}
\newblock \bibinfo{title}{Optical phase synchronization by injection of common broadband low-coherent light}.
\newblock \emph{\bibinfo{journal}{Phys. Rev. Lett.}} \textbf{\bibinfo{volume}{112}}, \bibinfo{pages}{204101} (\bibinfo{year}{2014}).

\bibitem{Macieszczak2016}
\bibinfo{author}{Macieszczak, K.}, \bibinfo{author}{Gu\c{t}\u{a}, M.}, \bibinfo{author}{Lesanovsky, I.} \& \bibinfo{author}{Garrahan, J.~P.}
\newblock \bibinfo{title}{Towards a theory of metastability in open quantum dynamics}.
\newblock \emph{\bibinfo{journal}{Phys. Rev. Lett.}} \textbf{\bibinfo{volume}{116}}, \bibinfo{pages}{240404} (\bibinfo{year}{2016}).

\bibitem{Note1}
\bibinfo{note}{In general, a more careful analysis is required that takes into account both the interplay between the eigenvalues and the eigenmodes to correctly estimate the relaxation time scales. Knowledge of the real part of the spectrum $\protect \mathrm {Re}[\Lambda _1],\protect \mathrm {Re}[\Lambda _m]$ and $\protect \mathrm {Re}[\Lambda _{m+1}]$ may no longer provide sufficient information, if for instance the coefficients $c_k, \ (k \ge m+1)$ are anomalously large and thus modify the decay rates at intermediate times \cite {Song2019,Mori2020,Haga2021,Lee2023}.}

\bibitem{Znidaric2015}
\bibinfo{author}{\v{Z}nidari\v{c}, M.}
\newblock \bibinfo{title}{Relaxation times of dissipative many-body quantum systems}.
\newblock \emph{\bibinfo{journal}{Phys. Rev. E}} \textbf{\bibinfo{volume}{92}}, \bibinfo{pages}{042143} (\bibinfo{year}{2015}).

\bibitem{Letscher2017}
\bibinfo{author}{Letscher, F.}, \bibinfo{author}{Thomas, O.}, \bibinfo{author}{Niederpr\"um, T.}, \bibinfo{author}{Fleischhauer, M.} \& \bibinfo{author}{Ott, H.}
\newblock \bibinfo{title}{Bistability versus metastability in driven dissipative rydberg gases}.
\newblock \emph{\bibinfo{journal}{Phys. Rev. X}} \textbf{\bibinfo{volume}{7}}, \bibinfo{pages}{021020} (\bibinfo{year}{2017}).

\bibitem{Minganti2018}
\bibinfo{author}{Minganti, F.}, \bibinfo{author}{Biella, A.}, \bibinfo{author}{Bartolo, N.} \& \bibinfo{author}{Ciuti, C.}
\newblock \bibinfo{title}{Spectral theory of liouvillians for dissipative phase transitions}.
\newblock \emph{\bibinfo{journal}{Phys. Rev. A}} \textbf{\bibinfo{volume}{98}}, \bibinfo{pages}{042118} (\bibinfo{year}{2018}).

\bibitem{Mori2020}
\bibinfo{author}{Mori, T.} \& \bibinfo{author}{Shirai, T.}
\newblock \bibinfo{title}{Resolving a discrepancy between {L}iouvillian gap and relaxation time in boundary-dissipated quantum many-body systems}.
\newblock \emph{\bibinfo{journal}{Phys. Rev. Lett.}} \textbf{\bibinfo{volume}{125}}, \bibinfo{pages}{230604} (\bibinfo{year}{2020}).

\bibitem{Haga2021}
\bibinfo{author}{Haga, T.}, \bibinfo{author}{Nakagawa, M.}, \bibinfo{author}{Hamazaki, R.} \& \bibinfo{author}{Ueda, M.}
\newblock \bibinfo{title}{Liouvillian skin effect: Slowing down of relaxation processes without gap closing}.
\newblock \emph{\bibinfo{journal}{Phys. Rev. Lett.}} \textbf{\bibinfo{volume}{127}}, \bibinfo{pages}{070402} (\bibinfo{year}{2021}).

\bibitem{Lee2023}
\bibinfo{author}{Lee, G.}, \bibinfo{author}{McDonald, A.} \& \bibinfo{author}{Clerk, A.}
\newblock \bibinfo{title}{Anomalously large relaxation times in dissipative lattice models beyond the non-{H}ermitian skin effect}.
\newblock \emph{\bibinfo{journal}{Phys. Rev. B}} \textbf{\bibinfo{volume}{108}}, \bibinfo{pages}{064311} (\bibinfo{year}{2023}).

\bibitem{Facchi2008}
\bibinfo{author}{Facchi, P.} \& \bibinfo{author}{Pascazio, S.}
\newblock \bibinfo{title}{Quantum {Z}eno dynamics: mathematical and physical aspects}.
\newblock \emph{\bibinfo{journal}{J. Phys. A: Math. Theor.}} \textbf{\bibinfo{volume}{41}}, \bibinfo{pages}{493001} (\bibinfo{year}{2008}).

\bibitem{Zanardi2014}
\bibinfo{author}{Zanardi, P.} \& \bibinfo{author}{Campos~Venuti, L.}
\newblock \bibinfo{title}{Coherent quantum dynamics in steady-state manifolds of strongly dissipative systems}.
\newblock \emph{\bibinfo{journal}{Phys. Rev. Lett.}} \textbf{\bibinfo{volume}{113}}, \bibinfo{pages}{240406} (\bibinfo{year}{2014}).

\bibitem{Popkov2018}
\bibinfo{author}{Popkov, V.}, \bibinfo{author}{Essink, S.}, \bibinfo{author}{Presilla, C.} \& \bibinfo{author}{Sch\"utz, G.}
\newblock \bibinfo{title}{Effective quantum {Z}eno dynamics in dissipative quantum systems}.
\newblock \emph{\bibinfo{journal}{Phys. Rev. A}} \textbf{\bibinfo{volume}{98}}, \bibinfo{pages}{052110} (\bibinfo{year}{2018}).

\bibitem{Burgarth2020}
\bibinfo{author}{Burgarth, D.}, \bibinfo{author}{Facchi, P.}, \bibinfo{author}{Nakazato, H.}, \bibinfo{author}{Pascazio, S.} \& \bibinfo{author}{Yuasa, K.}
\newblock \bibinfo{title}{Quantum {Z}eno {D}ynamics from {G}eneral {Q}uantum {O}perations}.
\newblock \emph{\bibinfo{journal}{{Quantum}}} \textbf{\bibinfo{volume}{4}}, \bibinfo{pages}{289} (\bibinfo{year}{2020}).

\bibitem{Popkov2021}
\bibinfo{author}{Popkov, V.} \& \bibinfo{author}{Presilla, C.}
\newblock \bibinfo{title}{Full spectrum of the liouvillian of open dissipative quantum systems in the zeno limit}.
\newblock \emph{\bibinfo{journal}{Phys. Rev. Lett.}} \textbf{\bibinfo{volume}{126}}, \bibinfo{pages}{190402} (\bibinfo{year}{2021}).

\bibitem{Breuer2016}
\bibinfo{author}{Breuer, H.-P.}, \bibinfo{author}{Laine, E.-M.}, \bibinfo{author}{Piilo, J.} \& \bibinfo{author}{Vacchini, B.}
\newblock \bibinfo{title}{Colloquium: Non-{M}arkovian dynamics in open quantum systems}.
\newblock \emph{\bibinfo{journal}{Rev. Mod. Phys.}} \textbf{\bibinfo{volume}{88}}, \bibinfo{pages}{021002} (\bibinfo{year}{2016}).

\bibitem{deVega2017}
\bibinfo{author}{de~Vega, I.} \& \bibinfo{author}{Alonso, D.}
\newblock \bibinfo{title}{Dynamics of non-{M}arkovian open quantum systems}.
\newblock \emph{\bibinfo{journal}{Rev. Mod. Phys.}} \textbf{\bibinfo{volume}{89}}, \bibinfo{pages}{015001} (\bibinfo{year}{2017}).

\bibitem{Karpat2021}
\bibinfo{author}{Karpat, G.}, \bibinfo{author}{Yal\c{c}{\i}nkaya, I.}, \bibinfo{author}{\c{C}akmak, B.~c.}, \bibinfo{author}{Giorgi, G.~L.} \& \bibinfo{author}{Zambrini, R.}
\newblock \bibinfo{title}{Synchronization and non-{M}arkovianity in open quantum systems}.
\newblock \emph{\bibinfo{journal}{Phys. Rev. A}} \textbf{\bibinfo{volume}{103}}, \bibinfo{pages}{062217} (\bibinfo{year}{2021}).

\bibitem{Zhou2021}
\bibinfo{author}{Zhou, K.-J.}, \bibinfo{author}{Zou, J.}, \bibinfo{author}{Xu, B.-M.}, \bibinfo{author}{Li, L.} \& \bibinfo{author}{Shao, B.}
\newblock \bibinfo{title}{Effect of non-{M}arkovianity on synchronization}.
\newblock \emph{\bibinfo{journal}{Comm. Theor. Phys.}} \textbf{\bibinfo{volume}{73}}, \bibinfo{pages}{105101} (\bibinfo{year}{2021}).

\bibitem{Wilczek2012}
\bibinfo{author}{Wilczek, F.}
\newblock \bibinfo{title}{Quantum time crystals}.
\newblock \emph{\bibinfo{journal}{Phys. Rev. Lett.}} \textbf{\bibinfo{volume}{109}}, \bibinfo{pages}{160401} (\bibinfo{year}{2012}).

\bibitem{Sacha2018}
\bibinfo{author}{Sacha, K.} \& \bibinfo{author}{Zakrzewski, J.}
\newblock \bibinfo{title}{Time crystals: a review}.
\newblock \emph{\bibinfo{journal}{Rep. Prog. Phys.}} \textbf{\bibinfo{volume}{81}}, \bibinfo{pages}{016401} (\bibinfo{year}{2017}).

\bibitem{Khemani2019}
\bibinfo{author}{Khemani, V.}, \bibinfo{author}{Moessner, R.} \& \bibinfo{author}{Sondhi, S.~L.}
\newblock \bibinfo{title}{A brief history of time crystals}.
\newblock \emph{\bibinfo{journal}{arXiv:1910.10745}}  (\bibinfo{year}{2019}).

\bibitem{Zaletel2023}
\bibinfo{author}{Zaletel, M.~P.} \emph{et~al.}
\newblock \bibinfo{title}{Colloquium: Quantum and classical discrete time crystals}.
\newblock \emph{\bibinfo{journal}{Rev. Mod. Phys.}} \textbf{\bibinfo{volume}{95}}, \bibinfo{pages}{031001} (\bibinfo{year}{2023}).

\bibitem{Iemini2018}
\bibinfo{author}{Iemini, F.} \emph{et~al.}
\newblock \bibinfo{title}{Boundary time crystals}.
\newblock \emph{\bibinfo{journal}{Phys. Rev. Lett.}} \textbf{\bibinfo{volume}{121}}, \bibinfo{pages}{035301} (\bibinfo{year}{2018}).

\bibitem{Hadjusek2022}
\bibinfo{author}{Hajdu\v{s}ek, M.}, \bibinfo{author}{Solanki, P.}, \bibinfo{author}{Fazio, R.} \& \bibinfo{author}{Vinjanampathy, S.}
\newblock \bibinfo{title}{Seeding crystallization in time}.
\newblock \emph{\bibinfo{journal}{Phys. Rev. Lett.}} \textbf{\bibinfo{volume}{128}}, \bibinfo{pages}{080603} (\bibinfo{year}{2022}).

\bibitem{Krishna2023}
\bibinfo{author}{Krishna, M.}, \bibinfo{author}{Solanki, P.}, \bibinfo{author}{Hajdu\v{s}ek, M.} \& \bibinfo{author}{Vinjanampathy, S.}
\newblock \bibinfo{title}{Measurement-induced continuous time crystals}.
\newblock \emph{\bibinfo{journal}{Phys. Rev. Lett.}} \textbf{\bibinfo{volume}{130}}, \bibinfo{pages}{150401} (\bibinfo{year}{2023}).

\bibitem{Solanki2024}
\bibinfo{author}{Solanki, P.}, \bibinfo{author}{Krishna, M.}, \bibinfo{author}{Hajdu\v{s}ek, M.}, \bibinfo{author}{Bruder, C.} \& \bibinfo{author}{Vinjanampathy, S.}
\newblock \bibinfo{title}{Exotic synchronization in continuous time crystals outside the symmetric subspace}.
\newblock \emph{\bibinfo{journal}{Phys. Rev. Lett.}} \textbf{\bibinfo{volume}{133}}, \bibinfo{pages}{260403} (\bibinfo{year}{2024}).

\bibitem{Russo2025}
\bibinfo{author}{Russo, F.} \& \bibinfo{author}{Pohl, T.}
\newblock \bibinfo{title}{Quantum dissipative continuous time crystals}.
\newblock \emph{\bibinfo{journal}{Phys. Rev. Lett.}} \textbf{\bibinfo{volume}{135}}, \bibinfo{pages}{110404} (\bibinfo{year}{2025}).

\bibitem{Buca2019}
\bibinfo{author}{Buca, B.}, \bibinfo{author}{Tindall, J.} \& \bibinfo{author}{Jaksch, D.}
\newblock \bibinfo{title}{Non-stationary coherent quantum many-body dynamics through dissipation}.
\newblock \emph{\bibinfo{journal}{Nature Comm.}} \textbf{\bibinfo{volume}{10}} (\bibinfo{year}{2019}).

\bibitem{Buonaiuto2021}
\bibinfo{author}{Buonaiuto, G.}, \bibinfo{author}{Carollo, F.}, \bibinfo{author}{Olmos, B.} \& \bibinfo{author}{Lesanovsky, I.}
\newblock \bibinfo{title}{Dynamical phases and quantum correlations in an emitter-waveguide system with feedback}.
\newblock \emph{\bibinfo{journal}{Phys. Rev. Lett.}} \textbf{\bibinfo{volume}{127}}, \bibinfo{pages}{133601} (\bibinfo{year}{2021}).

\bibitem{Solanki2023}
\bibinfo{author}{Solanki, P.}, \bibinfo{author}{Mehdi, F.~M.}, \bibinfo{author}{Hajdu\v{s}ek, M.} \& \bibinfo{author}{Vinjanampathy, S.}
\newblock \bibinfo{title}{Symmetries and synchronization blockade}.
\newblock \emph{\bibinfo{journal}{Phys. Rev. A}} \textbf{\bibinfo{volume}{108}}, \bibinfo{pages}{022216} (\bibinfo{year}{2023}).

\bibitem{Dai2026}
\bibinfo{author}{Dai, S.} \emph{et~al.}
\newblock \bibinfo{title}{Universal interaction-based manipulation of quantum synchronization in spin oscillator networks}.
\newblock \emph{\bibinfo{journal}{Phys. Rev. B}} \textbf{\bibinfo{volume}{113}}, \bibinfo{pages}{054306} (\bibinfo{year}{2026}).

\bibitem{Laskar2020}
\bibinfo{author}{Laskar, A.~W.} \emph{et~al.}
\newblock \bibinfo{title}{Observation of quantum phase synchronization in spin-1 atoms}.
\newblock \emph{\bibinfo{journal}{Phys. Rev. Lett.}} \textbf{\bibinfo{volume}{125}}, \bibinfo{pages}{013601} (\bibinfo{year}{2020}).

\bibitem{Ameri2015}
\bibinfo{author}{Ameri, V.} \emph{et~al.}
\newblock \bibinfo{title}{Mutual information as an order parameter for quantum synchronization}.
\newblock \emph{\bibinfo{journal}{Phys. Rev. A}} \textbf{\bibinfo{volume}{91}}, \bibinfo{pages}{012301} (\bibinfo{year}{2015}).

\bibitem{Mari2013}
\bibinfo{author}{Mari, A.}, \bibinfo{author}{Farace, A.}, \bibinfo{author}{Didier, N.}, \bibinfo{author}{Giovannetti, V.} \& \bibinfo{author}{Fazio, R.}
\newblock \bibinfo{title}{Measures of quantum synchronization in continuous variable systems}.
\newblock \emph{\bibinfo{journal}{Phys. Rev. Lett.}} \textbf{\bibinfo{volume}{111}}, \bibinfo{pages}{103605} (\bibinfo{year}{2013}).

\bibitem{Witthaut2017}
\bibinfo{author}{Witthaut, D.}, \bibinfo{author}{Wimberger, S.}, \bibinfo{author}{Burioni, R.} \& \bibinfo{author}{Timme, M.}
\newblock \bibinfo{title}{Classical synchronization indicates persistent entanglement in isolated quantum systems}.
\newblock \emph{\bibinfo{journal}{Nature Comm.}} \textbf{\bibinfo{volume}{8}} (\bibinfo{year}{2017}).

\bibitem{Manzano2013}
\bibinfo{author}{Manzano, G.}, \bibinfo{author}{Galve, F.}, \bibinfo{author}{Giorgi, G.~L.}, \bibinfo{author}{Hern{\'a}ndez-Garc{\'i}a, E.} \& \bibinfo{author}{Zambrini, R.}
\newblock \bibinfo{title}{Synchronization, quantum correlations and entanglement in oscillator networks}.
\newblock \emph{\bibinfo{journal}{Sci. Rep.}} \textbf{\bibinfo{volume}{3}}, \bibinfo{pages}{1439} (\bibinfo{year}{2013}).

\bibitem{Cabot2018}
\bibinfo{author}{Cabot, A.} \emph{et~al.}
\newblock \bibinfo{title}{Unveiling noiseless clusters in complex quantum networks}.
\newblock \emph{\bibinfo{journal}{npj Quant. Info.}} \textbf{\bibinfo{volume}{4}} (\bibinfo{year}{2018}).

\bibitem{Tindall2020}
\bibinfo{author}{Tindall, J.}, \bibinfo{author}{Sanchez~Munoz, C.}, \bibinfo{author}{Buca, B.} \& \bibinfo{author}{Jaksch, D.}
\newblock \bibinfo{title}{Quantum synchronisation enabled by dynamical symmetries and dissipation}.
\newblock \emph{\bibinfo{journal}{New J. Phys.}} \textbf{\bibinfo{volume}{22}}, \bibinfo{pages}{013026} (\bibinfo{year}{2020}).

\bibitem{Tong2025}
\bibinfo{author}{Liu, T.}, \bibinfo{author}{Garc\'{\i}a-\'Alvarez, L.} \& \bibinfo{author}{Tancredi, G.}
\newblock \bibinfo{title}{Quantum synchronization in one-dimensional topological systems}.
\newblock \emph{\bibinfo{journal}{Phys. Rev. Res.}} \textbf{\bibinfo{volume}{7}}, \bibinfo{pages}{L022064} (\bibinfo{year}{2025}).

\bibitem{Cakmak2026}
\bibinfo{author}{Cakmak, B.}, \bibinfo{author}{S\"umer, K.}, \bibinfo{author}{Campbell, S.} \& \bibinfo{author}{Karpat, G.}
\newblock \bibinfo{title}{Synchronization in a dissipative quantum many-body system}.
\newblock \emph{\bibinfo{journal}{arXiv:2604.18707}}  (\bibinfo{year}{2026}).

\bibitem{Jaseem2020}
\bibinfo{author}{Jaseem, N.} \emph{et~al.}
\newblock \bibinfo{title}{Generalized measure of quantum synchronization}.
\newblock \emph{\bibinfo{journal}{Phys. Rev. Res.}} \textbf{\bibinfo{volume}{2}}, \bibinfo{pages}{043287} (\bibinfo{year}{2020}).

\bibitem{Li2025}
\bibinfo{author}{Li, Y.} \emph{et~al.}
\newblock \bibinfo{title}{Experimental realization and synchronization of a quantum van der {P}ol oscillator}.
\newblock \emph{\bibinfo{journal}{Science Advances}} \textbf{\bibinfo{volume}{11}}, \bibinfo{pages}{eady5649} (\bibinfo{year}{2025}).

\bibitem{Koppenhofer2020}
\bibinfo{author}{Koppenh\"ofer, M.}, \bibinfo{author}{Bruder, C.} \& \bibinfo{author}{Roulet, A.}
\newblock \bibinfo{title}{Quantum synchronization on the {IBM Q} system}.
\newblock \emph{\bibinfo{journal}{Phys. Rev. Res.}} \textbf{\bibinfo{volume}{2}}, \bibinfo{pages}{023026} (\bibinfo{year}{2020}).

\bibitem{Tao2025}
\bibinfo{author}{Tao, Z.} \emph{et~al.}
\newblock \bibinfo{title}{Noise-induced quantum synchronization with entangled oscillations}.
\newblock \emph{\bibinfo{journal}{Nature Comm.}} \textbf{\bibinfo{volume}{16}}, \bibinfo{pages}{8457} (\bibinfo{year}{2025}).

\bibitem{Krithika2022}
\bibinfo{author}{Krithika, V.~R.}, \bibinfo{author}{Solanki, P.}, \bibinfo{author}{Vinjanampathy, S.} \& \bibinfo{author}{Mahesh, T.~S.}
\newblock \bibinfo{title}{Observation of quantum phase synchronization in a nuclear-spin system}.
\newblock \emph{\bibinfo{journal}{Phys. Rev. A}} \textbf{\bibinfo{volume}{105}}, \bibinfo{pages}{062206} (\bibinfo{year}{2022}).

\bibitem{Zhang2023}
\bibinfo{author}{Zhang, L.} \emph{et~al.}
\newblock \bibinfo{title}{Quantum synchronization of a single trapped-ion qubit}.
\newblock \emph{\bibinfo{journal}{Phys. Rev. Res.}} \textbf{\bibinfo{volume}{5}}, \bibinfo{pages}{033209} (\bibinfo{year}{2023}).

\bibitem{Liuvdp2025}
\bibinfo{author}{Liu, J.}, \bibinfo{author}{Wu, Q.}, \bibinfo{author}{Moore, J.~E.}, \bibinfo{author}{Haeffner, H.} \& \bibinfo{author}{W\"achtler, C.~W.}
\newblock \bibinfo{title}{Observation of synchronization between two quantum van der {P}ol oscillators in trapped ions}.
\newblock \emph{\bibinfo{journal}{arXiv:2509.18423}}  (\bibinfo{year}{2025}).

\bibitem{Wachtler2024}
\bibinfo{author}{W\"achtler, C.~W.} \& \bibinfo{author}{Moore, J.~E.}
\newblock \bibinfo{title}{Topological quantum synchronization of fractionalized spins}.
\newblock \emph{\bibinfo{journal}{Phys. Rev. Lett.}} \textbf{\bibinfo{volume}{132}}, \bibinfo{pages}{196601} (\bibinfo{year}{2024}).

\bibitem{li17}
\bibinfo{author}{Li, W.}, \bibinfo{author}{Li, C.} \& \bibinfo{author}{Song, H.}
\newblock \bibinfo{title}{Quantum synchronization and quantum state sharing in an irregular complex network}.
\newblock \emph{\bibinfo{journal}{Phys. Rev. E}} \textbf{\bibinfo{volume}{95}}, \bibinfo{pages}{022204} (\bibinfo{year}{2017}).

\bibitem{plj17}
\bibinfo{author}{Pljonkin, A.}, \bibinfo{author}{Rumyantsev, K.} \& \bibinfo{author}{Kumar~Singh, P.}
\newblock \bibinfo{title}{Synchronization in quantum key distribution systems}.
\newblock \emph{\bibinfo{journal}{Cryptography}} \textbf{\bibinfo{volume}{1}}, \bibinfo{pages}{18} (\bibinfo{year}{2017}).

\bibitem{Lohe2010}
\bibinfo{author}{Lohe, M.~A.}
\newblock \bibinfo{title}{Quantum synchronization over quantum networks}.
\newblock \emph{\bibinfo{journal}{J. Phys. A: Math. Theor.}} \textbf{\bibinfo{volume}{43}}, \bibinfo{pages}{465301} (\bibinfo{year}{2010}).

\bibitem{Nokkala2024}
\bibinfo{author}{Nokkala, J.}, \bibinfo{author}{Piilo, J.} \& \bibinfo{author}{Bianconi, G.}
\newblock \bibinfo{title}{Complex quantum networks: a topical review}.
\newblock \emph{\bibinfo{journal}{J. Phys. A: Math. Theor.}} \textbf{\bibinfo{volume}{57}}, \bibinfo{pages}{233001} (\bibinfo{year}{2024}).

\bibitem{vai25}
\bibinfo{author}{Vaidya, G.~M.}, \bibinfo{author}{J\"ager, S.~B.} \& \bibinfo{author}{Shankar, A.}
\newblock \bibinfo{title}{Quantum synchronization and dissipative quantum sensing}.
\newblock \emph{\bibinfo{journal}{Phys. Rev. A}} \textbf{\bibinfo{volume}{111}}, \bibinfo{pages}{012410} (\bibinfo{year}{2025}).

\bibitem{Song2019}
\bibinfo{author}{Song, F.}, \bibinfo{author}{Yao, S.} \& \bibinfo{author}{Wang, Z.}
\newblock \bibinfo{title}{Non-hermitian skin effect and chiral damping in open quantum systems}.
\newblock \emph{\bibinfo{journal}{Phys. Rev. Lett.}} \textbf{\bibinfo{volume}{123}}, \bibinfo{pages}{170401} (\bibinfo{year}{2019}).

\end{thebibliography}
\end{document}